\documentclass[twocolumn,aps,prb,amsmath,floatfix]{revtex4}
\usepackage{graphicx}
\begin{document}
\title{Pseudogap, non-Fermi-liquid behavior, and particle-hole asymmetry \\
               in  the 2D Hubbard model}
\author{Ansgar Liebsch$^1$ and Ning-Hua Tong$^2$} 
\affiliation{$^1$Institut f\"ur Festk\"orperforschung, 
             Forschungszentrum J\"ulich, 
             52425 J\"ulich, Germany \\
             $^2$Department of Physics, Renmin University of China, 
             100872 Beijing, China}
 \begin{abstract}
The effect of doping in the two-dimensional Hubbard model is studied
within finite temperature exact diagonalization combined with cluster 
dynamical mean field theory. By employing a mixed basis involving cluster
sites and bath molecular orbitals for the projection of the lattice 
Green's function onto $2\times2$ clusters, a considerably more accurate 
description of the low frequency properties of the self-energy is achieved 
than in a pure site picture.
The transition from Fermi-liquid to non-Fermi-liquid behavior for 
decreasing hole doping is studied as a function of Coulomb energy, 
next-nearest neighbor hopping, and temperature. In particular, the  
self-energy component $\Sigma_X$ associated with $X=(\pi,0)$ is shown 
to exhibit an onset of non-Fermi-liquid behavior as the hole doping 
decreases below a critical value $\delta_c$. The imaginary part of 
$\Sigma_X(\omega)$ then develops a collective mode above $E_F$,
which exhibits a distinct dispersion with doping. Accordingly, the 
real part of $\Sigma_X(\omega)$ has a positive slope above $E_F$, 
giving rise to an increasing particle-hole asymmetry as the system 
approaches the Mott transition. This behavior is consistent with the 
removal of spectral weight from electron states close to $E_F$ and 
the opening of a pseudogap which increases with decreasing doping. 
The phase diagram reveals that $\delta_c\approx 0.15\ldots0.20$ for 
various system parameters. For electron doping, the collective mode of 
$\Sigma_X(\omega)$ and the concomitant pseudogap are located below the 
Fermi energy which is consistent the removal of spectral weight from hole 
states just below $E_F$. The critical doping which marks the onset of 
non-Fermi-liquid behavior, is systematically smaller than for hole doping.     
\\
\mbox{\hskip1cm}  
electron systems 
\end{abstract}
\maketitle

\section{Introduction}

The nature of the metal insulator transition as a function of doping 
is one of the key issues in strongly correlated materials.
\cite{imada}
Experimental studies of many high-$T_c$ cuprates reveal a rich phase 
diagram, with conventional Fermi-liquid behavior in overdoped metals 
and an anomalous pseudogap phase in underdoped systems close to the 
Mott insulator.
One of the most intriguing and challenging aspects of the non-Fermi-liquid
phase is the observation of highly non-isotropic behavior in
momentum space. \cite{norman} 
Whereas along the nodal direction $\Gamma M$ well-defined quasiparticles
exist, in the vicinity of $X=(\pi,0)$ strong deviations from Fermi-liquid 
behavior occur. In particular, below a critical doping a pseudogap appears
which becomes more prominent close to the Mott insulator.
This transition from Fermi-liquid to non-Fermi-liquid properties has been 
widely investigated in recent years, and several theoretical models have
been proposed. 
\cite{pwa,wen,kivelson,varma,vojta,Chakravarty,honerkamp,lauchli,konik,%
yang1,tsvelik,sachdev,phillips}
          
Dynamical mean field theory\cite{dmft1,dmft2,dmft3,dmft4,dmft5,dmft} 
(DMFT) provides an elegant and 
successful framework for the description of the correlation induced 
transition from metallic to Mott insulating behavior.
\cite{reviews}
The local or single-site version of DMFT, however, focusses exclusively
on dynamical correlations which can give rise to spectral
weight transfer between low and high frequencies. To address the momentum
dependence of the self-energy, it is important to allow for
spatial fluctuations, at least on a short-range atomic scale.  
For this purpose, several approaches based on cluster extensions of 
DMFT \cite{hettler,lichtenstein,kotliar01,maier} as well as 
cluster perturbation theory\cite{senechal1} have been proposed.      
The general consensus that has emerged from many studies in this field
\cite{hettler2,jarrell2001,huscroft,moukouri,imai,maier2,onoda,%
kyung,potthoff,parcollet,capone,senechal2,civelli,choy,macridin06,%
tremblay,kyung1,kyung2,kyung3,capone2,stanescu,stanescu2,%
merino,macridin07,zhang,haule,park,macridin,senthil,civelli2,gull,balzer,%
ferrero,rubtsov,sakai,vidhya,yang2,werner,balzer2%
}
is that scattering processes are indeed much stronger close to $(\pi,0)$ 
and $(0,\pi)$ than in other regions of the Brillouin zone. 
Thus, Fermi-liquid behavior first breaks down in the antinodal direction 
and a pseudogap in the density of states opens up. In the nodal direction
between $(0,0)$ and $(\pi,\pi)$ Fermi-liquid behavior persists and 
well-defined quasiparticles can be identified.   

In the present work we use exact diagonalization\cite{ed} (ED) in 
combination with cellular DMFT\cite{kotliar01} (CDMFT) to investigate the 
two-dimensional Hubbard model on a square lattice for $2\times2$ clusters. 
For computational reasons, ED has previously been applied to study this 
model at $T=0$.\cite{civelli,kyung1,kyung2,stanescu2}
Here, we employ an extension to finite temperatures by making use of the 
Arnoldi algorithm\cite{arnoldi} which provides a highly efficient evaluation 
of excited states. Moreover, the cluster ED/DMFT is formulated in terms
of a mixed basis involving cluster sites and bath molecular orbitals 
which allows a very accurate projection of the lattice Green's function 
onto the $2\times2$ cluster.\cite{al2008} 
Thus, despite the use of only two bath levels per cluster orbital 
(12 levels in total), the spacing between excitation energies is very 
small, so that finite-size errors are greatly reduced, even at low
temperatures. As a result of these refinements, extrapolation from 
the Matsubara axis yields very accurate self-energies and Green's
functions at low real frequencies.    
The same approach has recently been used to evaluate
the phase diagram of the partially frustrated Hubbard model for 
triangular lattices.\cite{al2009}

The focus of this work is on the transition from Fermi-liquid to 
non-Fermi-liquid behavior for decreasing hole and electron doping. 
In particular, we study how this transition varies as a function of 
Coulomb energy, next-nearest neighbor hopping, and temperature. 
A systematic study of this variation is needed to explore the
phase diagram of the two-dimensional Hubbard model and has to our 
knowledge not been carried out before.  

The key quantity which exhibits the change from Fermi-liquid to 
non-Fermi-liquid behavior most clearly is the self-energy component 
$\Sigma_X$ associated with $X=(\pi,0)$. For hole doping 
$\delta\le 15\ldots20$~\%, spatial fluctuations within the cluster give 
rise to a collective mode in the imaginary part of $\Sigma_X(\omega)$ 
above $E_F$, in agreement with early work for $\delta=0.05$ by 
Jarrell {\it et al.} \cite{jarrell2001} based on quantum Monte Carlo (QMC) 
calculations within the Dynamical Cluster Approximation\cite{maier} (DCA).
The real part of $\Sigma_X(\omega)$ then exhibits a positive slope, 
implying removal of spectral weight from electron states close to $E_F$
and the opening of a pseudogap in the density of states. The evolution 
of this correlation-induced collective mode with decreasing doping leads 
to a widening of the pseudogap until it merges with the Mott gap at 
half-filling. In this region, the density of states acquires a very 
asymmetric shape. At large doping the Fermi level is located at a peak 
in the density of states, while for decreasing doping $E_F$ gradually 
shifts into the pseudogap, giving rise to a marked particle-hole  
asymmetry in the spectral distributions $A({\bf k},\omega)$ due to the 
reduced spectral weight above $E_F$. Moreover, with decreasing doping 
the pseudogap appears first along the antinodal direction before it 
opens across the entire Fermi surface. These results are in excellent 
correspondence with recent angle-resolved photoemission data for 
Bi$_2$Sr$_2$CaCu$_2$O$_{8+\delta}$ by Yang {\it et al.}\cite{yang3}   

The phase diagram shows that the change from Fermi-liquid to non-Fermi-liquid
behavior is remarkably stable, $\delta_c\approx 0.15\ldots0.20$, when 
system parameters, such as Coulomb energy, temperature, or second-neighbor 
hopping, are varied. For electron doping, the resonance of $\Sigma_X(\omega)$ 
is located below the Fermi energy, as expected for the removal of hole states 
just below $E_F$. The doping which defines the onset of non-Fermi-liquid 
behavior, is systematically smaller than for hole doping. Finally, 
the Mott transition induced by electron doping exhibits hysteresis 
behavior consistent with a first-order transition. In the case of hole 
doping, hysteresis behavior could not be identified at the temperatures
considered in this work.   
Thus, within the accuracy of our ED/CDMFT approach, this transition is 
either weakly first-order at very low temperatures or continuous.   
        
The outline of this paper is as follows. Section II presents the main
theoretical aspects of our finite $T$ cluster ED/DMFT approach.
Section III provides the results $2\times2$ clusters. In particular, 
we discuss the Mott transition, the non-Fermi-liquid properties, the
pseudogap, electron doping, the phase diagram, and the momentum
dependence. A summary is presented in Section IV.

\section{Cluster ED/DMFT in mixed site / orbital  basis} 

In this Section we outline the finite-temperature ED method in the
mixed site / molecular orbital basis which is employed as highly 
efficient and accurate impurity solver in the cluster DMFT.         
Let us consider the single-band Hubbard model for a two-dimensional
square lattice: 
\begin{equation}
H=-\sum_{\langle ij\rangle\sigma} t_{ij}( c^+_{i\sigma} c_{j\sigma} + {\rm H.c.}) 
                     + U \sum_i n_{i\uparrow} n_{i\downarrow} 
\end{equation}
where the sum in the first term extends up to second neighbors.
The band dispersion is given by 
$\epsilon({\bf k})= -2t[{\rm cos}(k_x)+{\rm cos}(k_y)]
            -4t'{\rm cos}(k_x){\rm cos}(k_y)$.
In order to approximately represent hole-doped cuprate systems, the 
nearest-neighbor hopping integral is defined as $t=0.25$ (band width $W=2$). 
The next-nearest-neighbor integral is mainly defined as $t'=-0.3t$, but 
$t'=0$ will also be considered. The local Coulomb interaction is taken 
to be $U=10 t= 2.5$ and  $U=6 t= 1.5$. Thus, at half-filling, the system 
is a Mott insulator. (For $t'=0$, QMC/DMFT calculations for 4-site clusters
\cite{park} yield $U_c\approx1.4\ldots1.5$, in agreement with ED/DMFT 
results for 2-site and 4-site clusters.\cite{al2008}
These values are consistent with recent QMC/DCA calculations for 8-site 
clusters\cite{werner} which give $U_c\approx1.4\ldots1.6$.)

Within CDMFT \cite{kotliar01} 
the interacting lattice Green's function in the cluster site basis is given  by
\begin{equation}
     G_{ij}(i\omega_n) = \sum_{\bf k} \left[ i\omega_n + \mu - t({\bf k})- 
                   \Sigma(i\omega_n)\right]^{-1}_{ij} 
\label{G}
\end{equation}
where the ${\bf k}$ sum extends over the reduced Brillouin Zone, 
$\omega_n=(2n+1)\pi T$ are Matsubara frequencies and $\mu$ is the chemical
potential. $t({\bf k})$ denotes the hopping matrix for the superlattice and 
$\Sigma_{ij}(i\omega_n)$ represents the cluster self-energy matrix.
The lattice constant is taken to be $a=1$ and site labels refer to 
$1\equiv(0,0)$, $2\equiv(1,0)$, $3\equiv(0,1)$, and $4\equiv(1,1)$.
In this geometry, all diagonal elements of the symmetric matrix $G_{ij}$ 
are identical and there are only two independent off-diagonal elements: 
$G_{12}=G_{13}=G_{24}=G_{34}$ and $G_{14}=G_{23}$.
By definition, both the lattice Green's function $G_{ij}$ and 
self-energy $\Sigma_{ij}$ have continuous spectral distributions at real 
$\omega$.  Only the paramagnetic phase will be considered here.

It is useful to transform the site basis into a molecular orbital basis 
in which the Green's function and self-energy become diagonal. 
The orbitals are defined as:\\
        $\phi_1=(|1\rangle+|2\rangle+|3\rangle+|4\rangle)/2$, 
        $\phi_2=(|1\rangle-|2\rangle-|3\rangle+|4\rangle)/2$,         
        $\phi_3=(|1\rangle+|2\rangle-|3\rangle-|4\rangle)/2$, 
        $\phi_4=(|1\rangle-|2\rangle+|3\rangle-|4\rangle)/2$.
We refer to these orbitals as $\Gamma$, $M$ and $X$, respectively,
where $X$ is doubly degenerate.
The Green's function elements in this basis will be denoted as 
$ G_{m}(i\omega_n)$, where 
\begin{eqnarray}
 G_\Gamma &\equiv& G_1=G_{11}+ 2 G_{12}+ G_{14} \nonumber \\ 
 G_M      &\equiv& G_2=G_{11}- 2 G_{12}+ G_{14} \\
 G_X      &\equiv& G_3=G_4 =G_{11}-G_{14}. \nonumber  
\label{Gk}
\end{eqnarray}
An analogous notation is used for the self-energy. Similar diagonal 
representations of $G$ and $\Sigma$ have been used in several previous 
works.\cite{jarrell2001,civelli,haule,park,koch,al2008,gull,balzer}

\begin{figure} 
\begin{center}
\includegraphics[width=4.5cm,height=6.5cm,angle=-90]{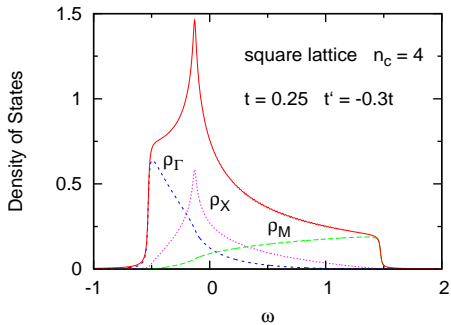}
\end{center}
\caption{(Color online)
Total density of states $\rho(\omega)$ and molecular orbital components 
$\rho_m(\omega)$ for four-site clusters of square lattice. 
For clarity, the molecular orbital components are divided by $n_c=4$.  
$\omega=0$ defines the Fermi energy for half-filling. At 14~\% hole doping
the van Hove singularity is shifted from $\omega=-0.13$ to $-0.09$.  
}\label{dos}\end{figure}

Figure~\ref{dos} illustrates the uncorrelated density of states components 
in the molecular orbital basis, where 
$\rho_m(\omega) = -\frac{1}{\pi}\,{\rm Im}\, G_m(\omega)$  for $\Sigma=0$,
and we denote $\rho_{\Gamma}=\rho_1$, $\rho_M=\rho_2$, $\rho_X=\rho_{3,4}$.   
The average or local density is $\rho_{av}=(\rho_{\Gamma}+\rho_M+2\rho_X)/4$.
Note that all molecular orbital densities extend across the entire band 
width. Nevertheless, only $\rho_X$ contains the van Hove singularity, 
while $\rho_\Gamma$ and $\rho_M$ are roughly representative of the spectral 
weight near ${\bf k}=(0,0)$ and ${\bf k}=(\pi,\pi)$, respectively.
Hole doping shifts the van Hove singularity towards $E_F$, 
whereas electron doping moves this singularity away from $E_F$.

A central feature of DMFT is that, to avoid double-counting of Coulomb 
interactions in the quantum impurity calculation, the self-energy must be
removed from the small cluster in which correlations are treated explicitly. 
This removal yields the Green's function 
\begin{equation}
         G_0(i\omega_n) = [G(i\omega_n)^{-1} + \Sigma(i\omega_n)]^{-1} , 
           \label{G0}
\end{equation}
which is also diagonal in the molecular orbital basis.

For the purpose of perfoming the ED calculation we now project the diagonal 
components of $G_0(i\omega_n)$ onto those of a larger cluster consisting of 
$n_c=4$ impurity levels and $n_b=8$ bath levels. The total number of levels 
is $n_s=n_c+n_b=12$. Thus,
\begin{eqnarray}
 G_{0,m}(i\omega_n) &\approx&  G^{cl}_{0,m}(i\omega_n) \nonumber\\
    &=&    \left( i\omega_n + \mu -\epsilon_m -
 \sum_{k=5}^{12} \frac{\vert V_{mk}\vert^2}{i\omega_n - \epsilon_k}\right)^{-1}
   \label{G0m}
\end{eqnarray}
where $\epsilon_m$ denote the molecular orbital levels, $\epsilon_k$ 
the bath levels, and $V_{mk}$ the hybridization matrix elements. The 
incorporation of the impurity level $\epsilon_m$ ensures a much better 
fit of $G_{0,m}(i\omega_n)$ than by projecting only onto bath orbitals. 

Assuming independent baths for the cluster orbitals, each component 
$G_{0,m}(i\omega_n)$ is fitted using five parameters: one impurity level 
$\epsilon_m$, two bath levels $\epsilon_k$ and two hopping integrals
$V_{mk}$. For instance, orbital 1 couples to bath levels 5 and 9, orbital 
2 to bath levels 6 and 10, etc. For the three independent cluster Green's 
functions, we therefore use a total of 15 fit parameters to represent 
$G_{0}(i\omega_n)$. This procedure provides a considerably more flexible 
projection than within a pure site basis. Since for symmetry reasons all 
sites are equivalent one would have in this case only four parameters 
(without including a level at the cluster sites). Thus, the molecular orbital 
basis allows for 11 additional cross hybridization terms as well as internal 
cluster couplings (see below). In addition, it is much more reliable to fit 
the three independent molecular orbital components $G_{0,m}(i\omega_n)$ than 
a non-diagonal site matrix $G_{0,ij}(i\omega_n)$ with only 4 parameters. 

\begin{figure}[t]  
\begin{center}
\includegraphics[width=4.5cm,height=6.5cm,angle=-90]{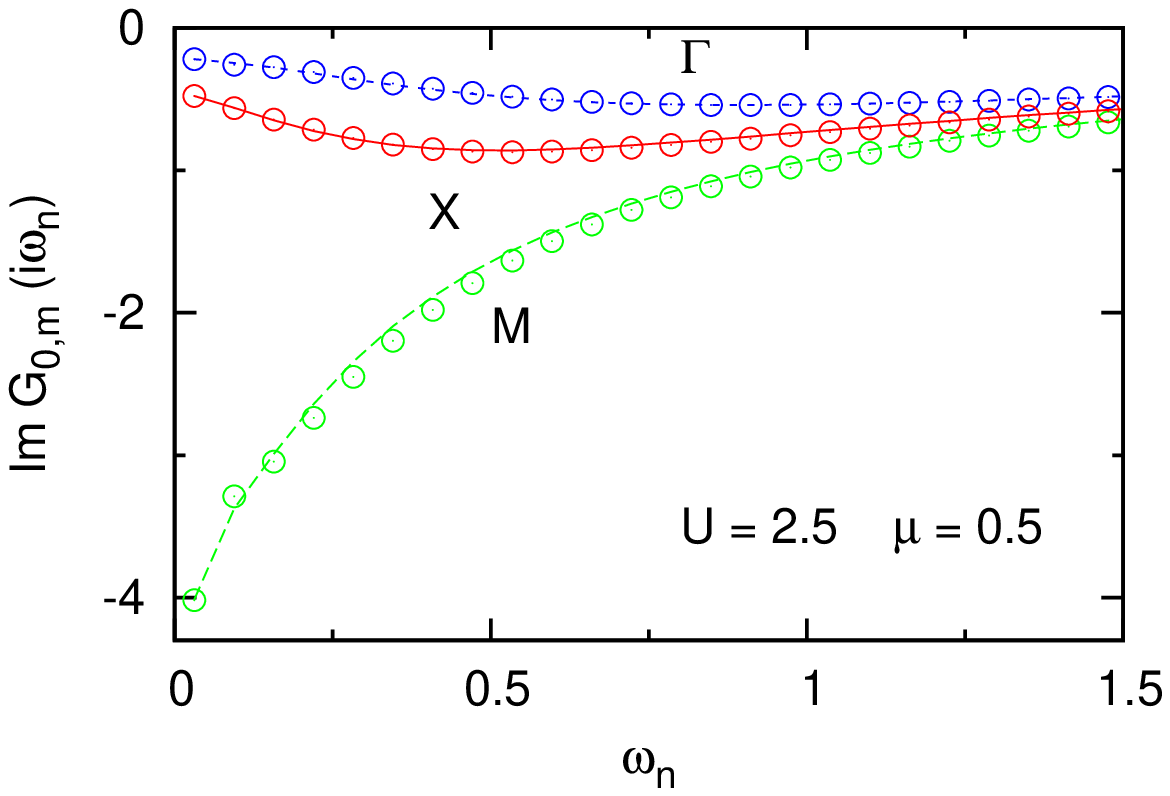}
\includegraphics[width=4.5cm,height=6.5cm,angle=-90]{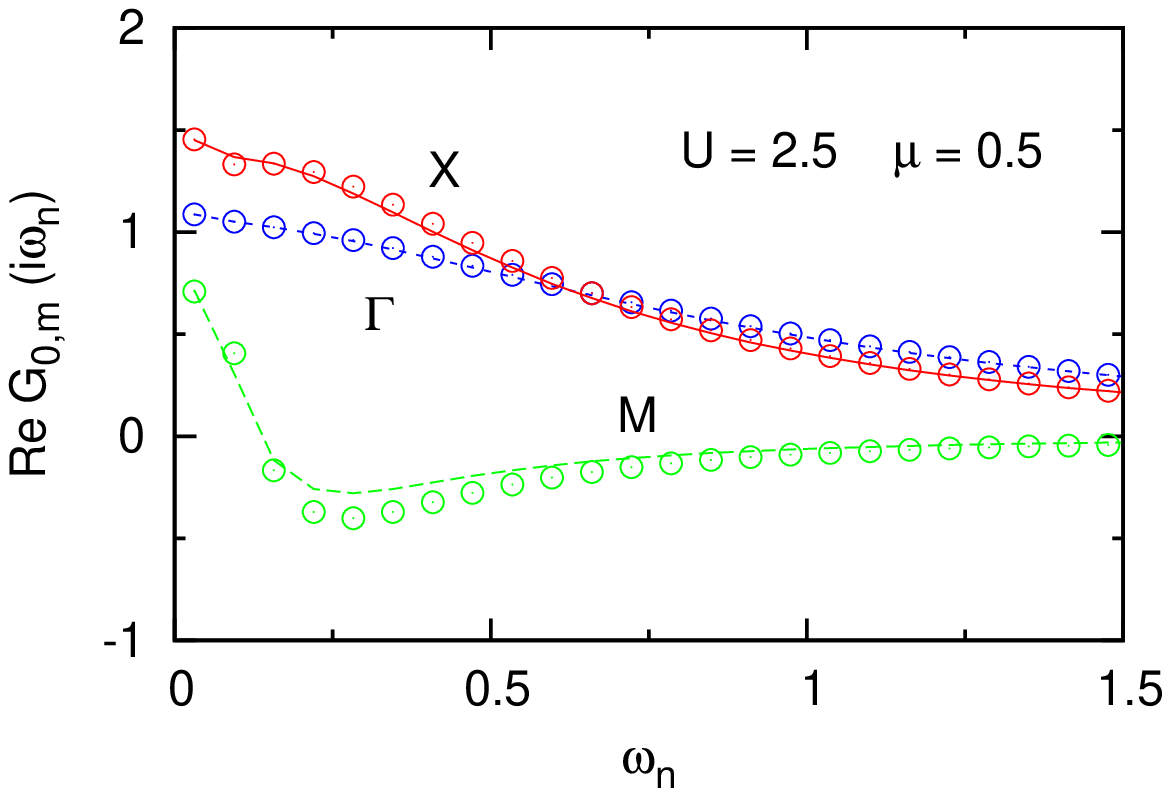}
\end{center} 
\caption{(Color online)
Projection of lattice Green's function components $G_{0,m}(i\omega_n)$ 
onto cluster consisting of four impurity levels and eight bath levels, 
for $U=2.5$, $\mu=0.5$, $T=0.01$; $t'=-0.3t$. 
Upper panel: Im\,$G_{0,m}$, lower panel: Re\,$G_{0,m}$. 
Continuous curves: diagonal elements of lattice Green's function, Eq.~(\ref{G0});
circles: approximate expression, right-hand side of Eq.~(\ref{G0m}). 
}\label{g0}\end{figure}

Figure~\ref{g0} illustrates the projection of the lattice Green's function  
$G_{0}(i\omega_n)$ onto the cluster for $U=2.5$ and $\mu=0.5$, which
corresponds to about $\delta=0.08$ hole doping. Projections of similar
quality are achieved at other Coulomb energies and chemical potentials.

The evaluation of the finite temperature interacting cluster Green's 
function could in principle also be carried out in the molecular orbital 
basis. The Coulomb interaction then becomes a matrix containing many 
inter-orbital components. This step can be circumvented by using a mixed 
basis consisting of cluster sites $i$ and bath molecular orbitals $k$. 
Thus, the diagonal $8\times8$ subblock $h_b=(\epsilon_k\delta_{kk'})$ 
representing the bath levels  
remains unchanged, but the diagonal $4\times4$ cluster molecular orbital 
submatrix now becomes nondiagonal in the cluster site basis. 
The transformation between sites $i$ and orbitals $m$ is given by  
\begin{eqnarray}
     T_{im} &=& 0.5 \left( \begin{array}{rrrr}
                               1 &  1 &  1 &  1  \\
                               1 & -1 & -1 &  1  \\
                               1 & -1 &  1 & -1  \\
                               1 &  1 & -1 & -1  \\
                                  \end{array} \right)  . \label{T}
\end{eqnarray}
In this mixed basis, the site subblock of the cluster Hamiltonian becomes
\begin{eqnarray}
      h_c &=& \left( \begin{array}{llll}
                    \epsilon & \tau     & \tau     & \tau'     \\
                    \tau     & \epsilon & \tau'    & \tau      \\
                    \tau     & \tau'    & \epsilon & \tau      \\  
                    \tau'    & \tau     & \tau     & \epsilon  \\  
                                             \end{array} \right) 
         \end{eqnarray}
with $\epsilon=(\epsilon_1+\epsilon_2+2\epsilon_3)/4$, 
     $\tau    =(\epsilon_1+\epsilon_2-2\epsilon_3)/4$, and 
     $\tau'   =(\epsilon_1-\epsilon_2)/2            $.     
Note that the hopping elements $t$ and $t'$ of the original lattice 
Hamiltonian do not appear since they are effectively absorbed into 
$\tau$ and $\tau'$ via the molecular orbital cluster levels $\epsilon_m$ 
which are adjusted to fit $G_{0,m}(i\omega_n)$. Evidently, the procedure
outlined above not only includes hopping between cluster and bath. 
It also introduces three new parameters within the $2\times2$ cluster:
$\epsilon$, $\tau$, and $\tau'$.      
In the mixed basis, the hybridization matrix elements $V_{mk}$ 
between cluster and bath molecular orbitals introduced in Eq.~(\ref{G0m}) 
are transformed to new hybridization matrix elements between cluster sites 
$i$ and bath orbitals $k$. They are given by
\begin{equation}
       V'_{ik} = (T V)_{ik} = \sum_m T_{im} V_{mk}\ .
\end{equation}
Thus, the upper right $4\times8$ submatrix containing the cluster / bath
hybridization matrix elements is transformed from  
\begin{eqnarray}
&&  \left( \begin{array}{llllllll}
     V_5 & 0 & 0 & 0 & V_9 & 0 & 0 & 0     \\
     0 & V_6 & 0 & 0 & 0 & V_{10} & 0 & 0     \\
     0 & 0 & V_7 & 0 & 0 & 0 & V_{11} & 0     \\
     0 & 0 & 0 & V_8 & 0 & 0 & 0 & V_{12}     \\
                                             \end{array} \right)  \ \ \ \ 
         \end{eqnarray}
to 
\begin{eqnarray}
&&  \left( \begin{array}{rrrrrrrr}
     V_5 & V_6 & V_7 & V_8 &\ V_9 & V_{10} & V_{11} & V_{12}     \\
     V_5 &-V_6 &-V_7 & V_8 &\ V_9 &-V_{10} &-V_{11} & V_{12}     \\
     V_5 &-V_6 & V_7 &-V_8 &\ V_9 &-V_{10} & V_{11} &-V_{12}     \\
     V_5 & V_6 &-V_7 &-V_8 &\ V_9 & V_{10} &-V_{11} &-V_{12}     \\
                                             \end{array} \right)  \ \ \ \ \ \ 
         \end{eqnarray}
The single-particle part of the cluster Hamiltonian then reads
\begin{eqnarray}
      h_0 &=& \left( \begin{array}{ll}
                    h_c      & V'    \\
                    V'^t     & h_b   \\   \end{array} \right) . 
\end{eqnarray}

Adding the onsite Coulomb interactions to this hamiltonian, the non-diagonal 
interacting cluster Green's function at finite $T$ can be derived from the 
expression\cite{perroni,luca}
\begin{eqnarray}
 G^{cl}_{ij}(i\omega_n) &=& \frac{1}{Z} \sum_{\nu\mu}\,e^{-\beta E_\nu}\, 
          \Big(\frac{\langle\nu\vert c_{i\sigma}  \vert\mu\rangle 
                     \langle\mu\vert c_{j\sigma}^+\vert\nu\rangle}
                                  {E_\nu - E_\mu + i\omega_n}            \nonumber\\
       &&\hskip9mm + \ \ \frac{\langle\nu\vert c_{i\sigma}^+\vert\mu\rangle 
                               \langle\mu\vert c_{j\sigma}  \vert\nu\rangle}
                                  {E_\mu - E_\nu + i\omega_n} \Big)   
     \label{Gcl}
\end{eqnarray}
where $E_\nu$ and $|\nu \rangle$  denote the eigenvalues and eigenvectors of 
the Hamiltonian, $\beta=1/T$ and $Z=\sum_\nu {\rm exp}(-\beta E_\nu)$ is the 
partition function. At low temperatures only a small number of excited states 
in a few spin sectors contributes to $G^{cl}_{ij}$. They can be efficiently 
evaluated using the Arnoldi algorithm.\cite{arnoldi} The excited state Green's 
functions are computed using the Lanczos procedure. Further details are provided 
in Ref.\cite{perroni}. The non-diagonal elements of $G^{cl}_{ij}$ are derived 
by first evaluating the diagonal components $G^{cl}_{ii}$ and then using the 
relation
\begin{equation}
     G^{cl}_{(i+j)(i+j)} = G^{cl}_{ii}+G^{cl}_{ij}+G^{cl}_{ji}+ G^{cl}_{jj}.
\end{equation}
Since $G^{cl}_{ij}=G^{cl}_{ji}$, this yields: 
\begin{equation}
     G^{cl}_{ij} = \frac{1}{2}[G^{cl}_{(i+j)(i+j)} - G^{cl}_{ii} - G^{cl}_{jj}].
\end{equation}
The interacting cluster Green's function $G^{cl}_{ij}$ satisfies the same 
symmetry properties as $G_{ij}$ and $G_{0,ij}$. It may therefore also be 
diagonalized, yielding cluster molecular orbital components $G^{cl}_{m}$.
The cluster molecular orbital self-energies can then be defined by an 
expression analogous to Eqs.~(\ref{G0}):
\begin{equation}
\Sigma^{cl}_{m}(i\omega_n) = 1/G^{cl}_{0,m}(i\omega_n)-1/G^{cl}_{m}(i\omega_n) .
\label{Scl}
\end{equation} 
At real $\omega$, these cluster self-energy components, just like $G^{cl}_{0,m}$ 
and $G^{cl}_{m}$, have discrete spectral distributions.   

The key assumption in DMFT is now that the impurity cluster self-energy is 
a physically reasonable representation of the lattice self-energy. Thus, 
\begin{equation}
     \Sigma_{m}(i\omega_n) \approx \Sigma^{cl}_{m}(i\omega_n) ,  
\label{S}
\end{equation}
where, at real frequencies, $\Sigma_{m}$ is continuous.

In the next iteration step, these diagonal self-energy components are 
used as input in the lattice Green's function Eq.~(\ref{G}), which in the molecular
orbital basis is given by 
\begin{equation}
   G_{m}(i\omega_n) = \sum_{\bf k} \left[ i\omega_n + \mu - T t({\bf k}) T^{-1} 
                 -  \Sigma(i\omega_n)\right]^{-1}_{mm} 
\label{Gm}
\end{equation}
where $T$ is the transformation defined in  Eq.~(\ref{T}). Thus, except for 
the diagonalization which is carried out in the mixed site / molecular orbital 
basis, all other steps of the calculational procedure are performed in the 
diagonal orbital basis. Note that $T t({\bf k}) T^{-1}$ is not diagonal at 
general ${\bf k}$ points. As a result, all orbital components of 
$\Sigma(i\omega_n)$ contribute to each $G_{m}(i\omega_n)$. This feature of
CDMFT differs from DCA where one has a one-to-one relation between 
 $\Sigma_m(i\omega_n)$ and  $G_{m}(i\omega_n)$:\cite{maier}  
\begin{equation}
G^{\rm DCA}_{m}(i\omega_n) = \sum_{{\bf k}_m} \left[ i\omega_n + \mu - \epsilon({\bf k}) 
                 -  \Sigma_m(i\omega_n)\right]^{-1} 
\label{Gdca}
\end{equation}
where ${\bf k}_m$ labels the $m^{th}$ patch of the Brillouin zone.

The largest spin sector for $n_s=12$ is $n_\uparrow=n_\downarrow=6$ with 
dimension $N=853776$. The interacting cluster Hamiltonian matrix $h$ is 
extremely sparse, so that only about 20 non-zero matrix elements per row 
need to be stored. Since the Arnoldi algorithm requires only operations of 
the type $h\, u = v$, where $u,\ v$ are vectors of dimension $N$, the procedure     
outlined above can easily be parallelized.
At temperatures of the order of $T=0.005\ldots0.02$, one iteration takes 
about 15 to 60 min on 8 processors. Except near the Mott transition, 5 to 10
iterations are usually required to achieve self-consistency. 

We conclude this section by pointing out that, once iteration to self-consistency 
has been carried out, a periodic lattice Green's function may be constructed from 
the cluster components in Eq.~(\ref{G}) by using the superposition:\cite{parcollet}
\begin{eqnarray}
G({\bf k},i\omega_n)&=&\frac{1}{4}\sum_{ij=1}^4 e^{i{\bf k}\cdot({\bf R}_i-{\bf R}_j)} 
           G_{ij}(i\omega_n)  \nonumber\\   
&=& G_{11}(i\omega_n) +G_{12}(i\omega_n)[{\rm cos}(k_x)+{\rm cos}(k_y)]\nonumber \\
          && \ \ \ \ + \ G_{14}(i\omega_n){\rm cos}(k_x){\rm cos}(k_y)
\label{Glat}
\end{eqnarray}
At high-symmetry points, this definition coincides with the diagonal elements
introduced in Eq.~(\ref{Gk}). Thus,    
      $ G_\Gamma(i\omega_n) = G((0  ,  0),i\omega_n)$,
      $ G_M     (i\omega_n) = G((\pi,\pi),i\omega_n)$, and  
      $ G_X     (i\omega_n) = G((\pi,  0),i\omega_n)= G((0,\pi),i\omega_n)$.
At ${\bf k}=(\pi/2,\pi/2)$, $G$ coincides with the onsite Green's function 
$G_{11}=(G_\Gamma+G_M+2G_X)/4$.

\section{Results and Discussion}
\subsection{Mott Transition}

\begin{figure}[t]  
\begin{center}
\includegraphics[width=4.5cm,height=6.5cm,angle=-90]{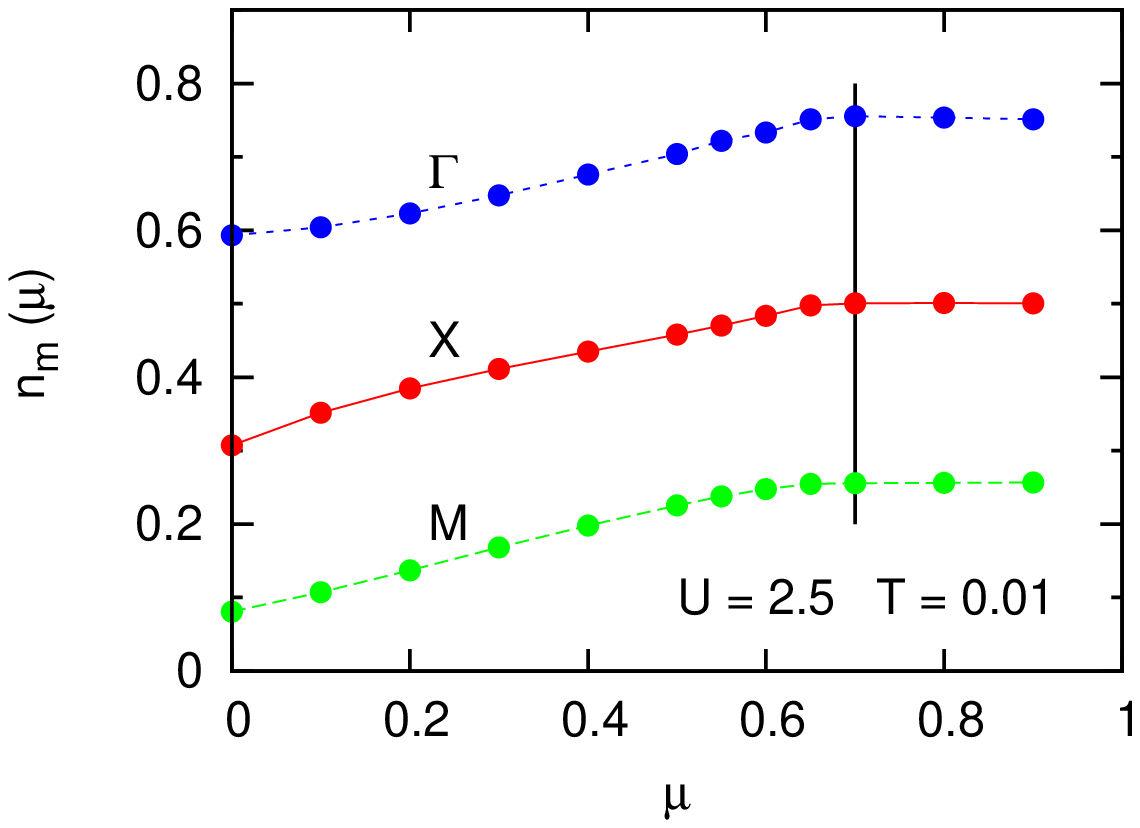}
\includegraphics[width=4.5cm,height=6.5cm,angle=-90]{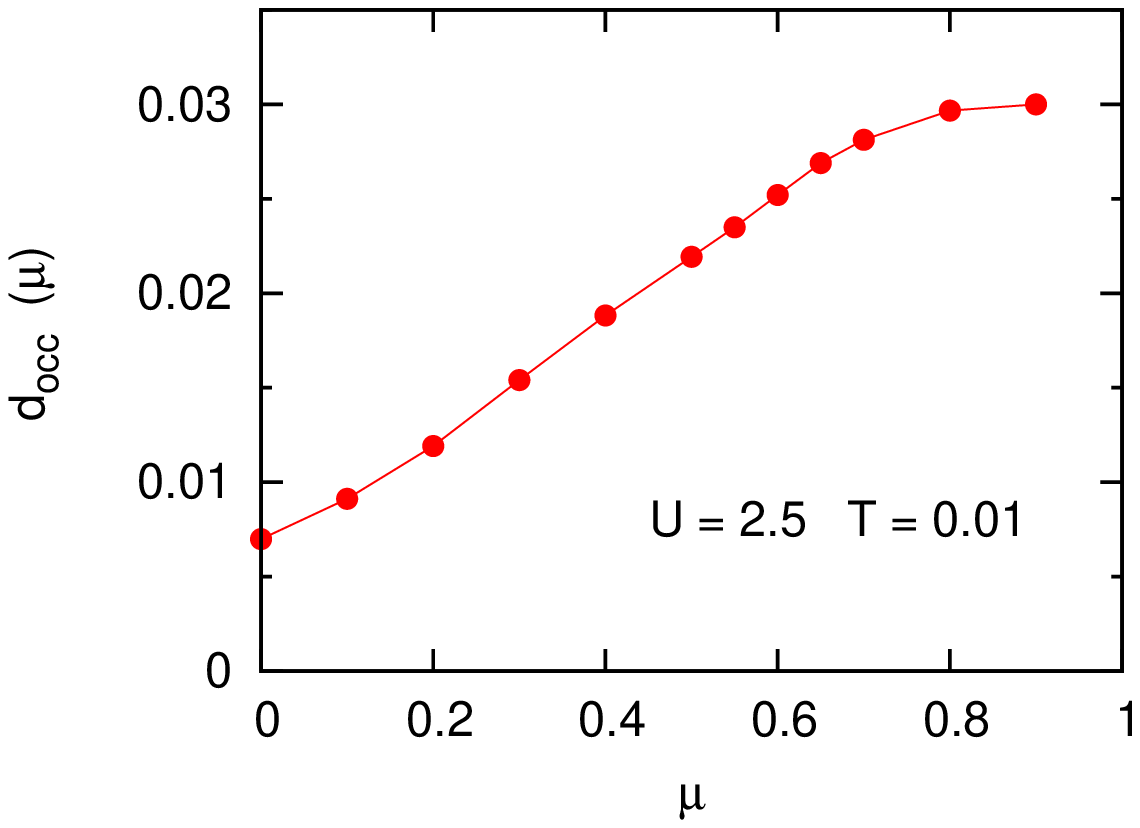}
\end{center}
\caption{(Color online) Upper panel:
Occupancies of cluster molecular orbitals (per spin) as functions of 
chemical potential $\mu$, for $U=2.5$, $T=0.01$, $t'=-0.3t$. The Mott transition 
occurs at about $\mu\approx 0.7$, indicated by the vertical bar, where 
$n_X\rightarrow0.5$, $n_\Gamma\rightarrow0.75$, and $n_M\rightarrow0.25$. 
Lower panel: Average double occupancy per site as a function of $\mu$.  
}\label{nvsmu}\end{figure}

Figure~\ref{nvsmu} shows the occupancies of the cluster molecular orbitals 
$\Gamma$, $M$ and $X$ as functions of chemical potential. 
The average occupancy per site (both spins) is 
$n=(n_\Gamma+n_M +2n_X)/2=1-\delta$, where $\delta$ 
is the hole doping. As revealed 
by the spectral distributions discussed below, the Mott transition
occurs at $\mu\approx 0.7$, where the $X$ orbital becomes half-filled,
whereas $n_\Gamma$ and $n_M$ approach $0.25$ and $0.75$, respectively.
Thus, all three orbitals take part in the transition. This result is 
consistent with previous ED/DMFT calculations\cite{al2008} for 2-site 
and 4-site clusters in the limit $t'=0$, and with recent QMC results
\cite{ferrero} for a minimal 2-site cluster DCA version,
where hole doping takes place at about the 
same rate for both inner and outer regions of the Brillouin zone.
These trends differ, however, from results for an 8-site continuous
time (CT) QMC/DCA calculation\cite{werner} 
which reveals initial doping primarily along the nodal direction, 
while near $X$ the occupancy for small $\delta$ remains at the same 
value as in the Mott insulator. Evidently, 2-site and 4-site cluster 
DMFT approaches do not provide sufficient momentum resolution to 
allow for ${\bf k}$-dependent doping.     

The lower panel of Fig.~3 shows the average double occupancy per site.
We have calculated these occupancies both for increasing and decreasing
chemical potential without encountering hysteresis behavior for $T\ge0.005$.  
The Mott transition induced by hole doping is therefore weakly first-order 
at even lower temperatures, or continuous. 
This result differs from the case of electron doping discussed farther 
below, where $n_m(\mu)$ as well as $d_{occ}(\mu)$ readily show hysteresis. 

\begin{figure}[t]  
\begin{center}
\includegraphics[width=4.5cm,height=6.5cm,angle=-90]{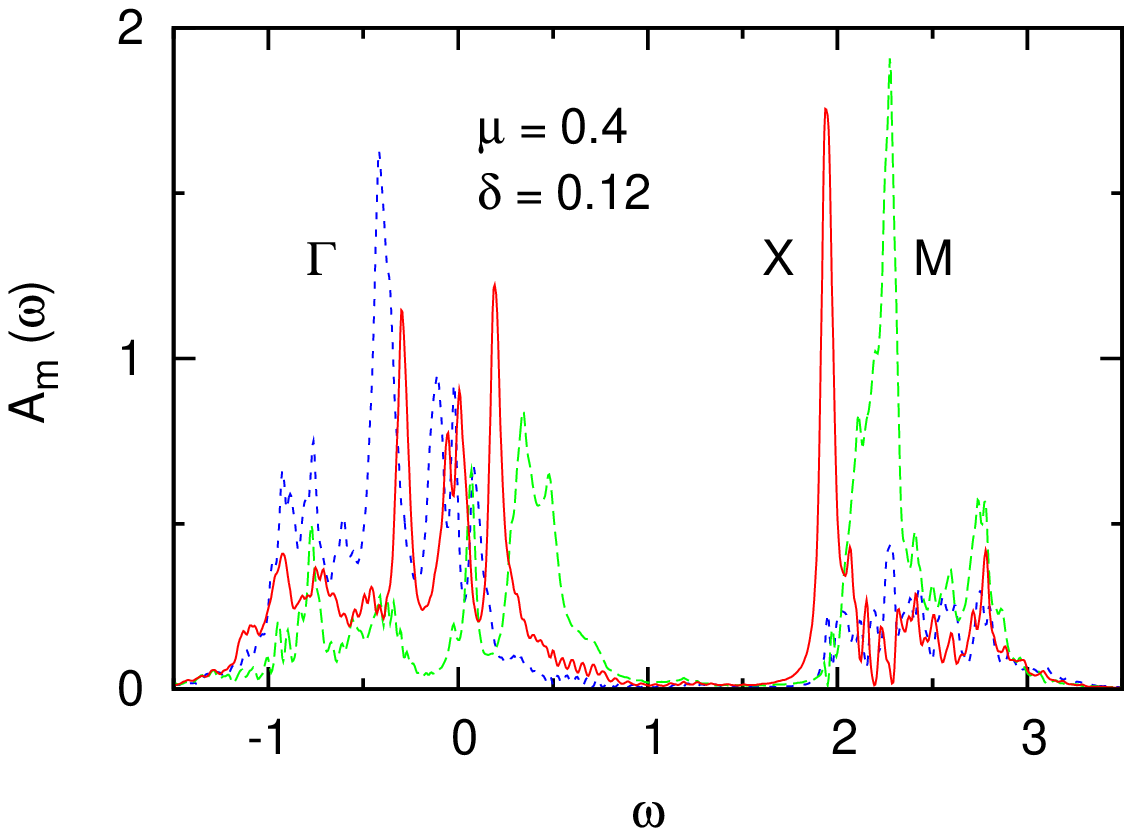}
\includegraphics[width=4.5cm,height=6.5cm,angle=-90]{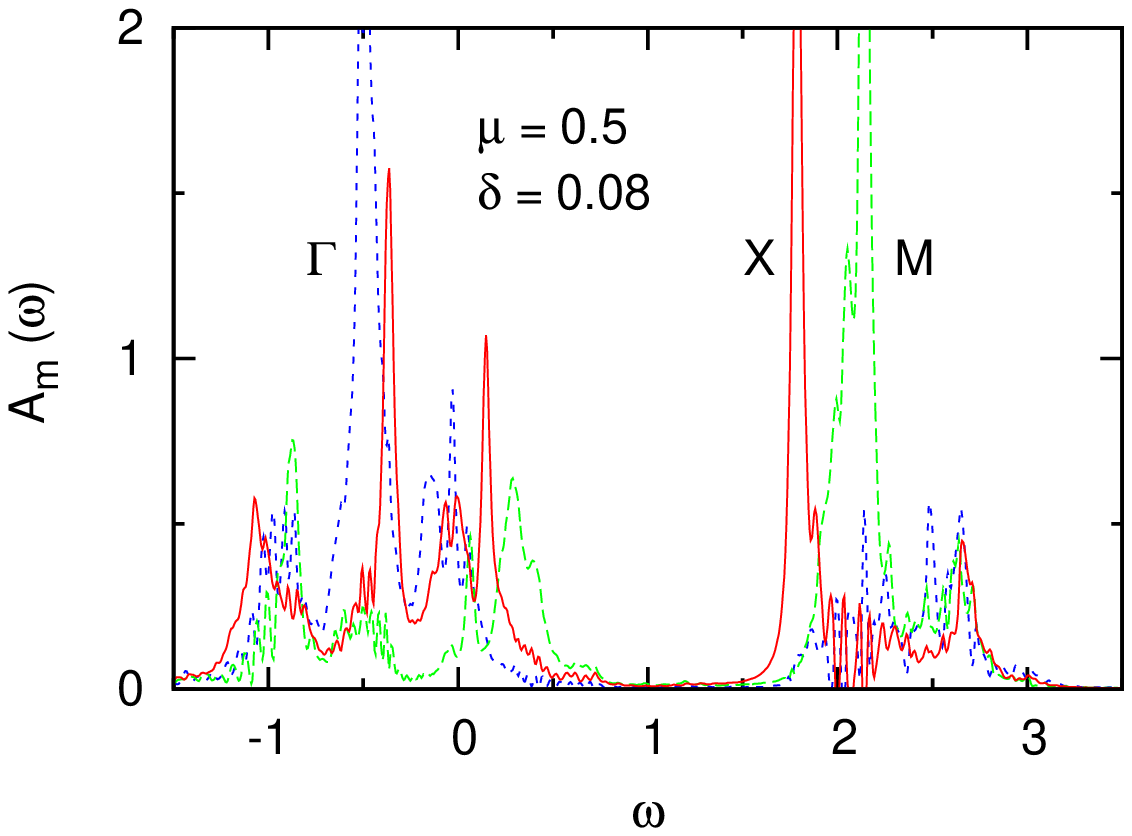}
\includegraphics[width=4.5cm,height=6.5cm,angle=-90]{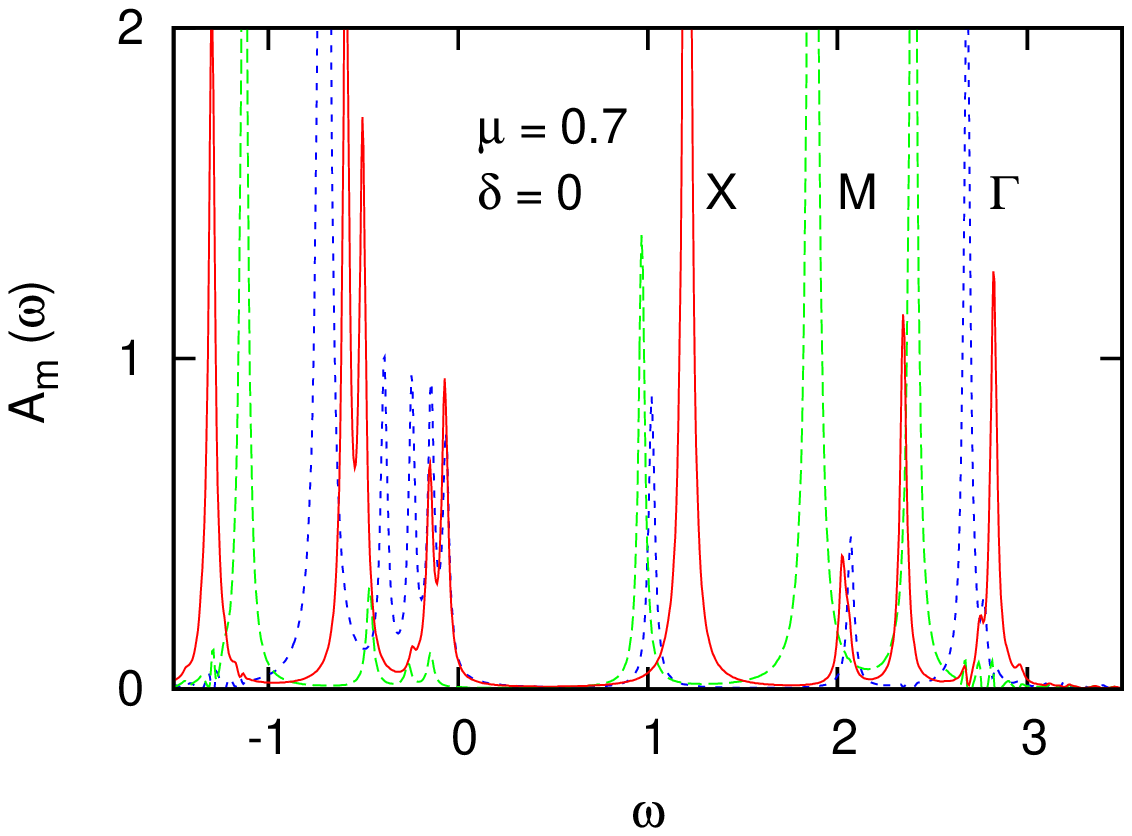}
\end{center}
\caption{(Color online)
Cluster spectral distributions for various chemical potentials corresponding
to hole doping $\delta=0.12$, $\delta=0.08$, and  $\delta=0$; $U=2.5$, 
$T=0.01$; broadening $\gamma=0.02$. 
Solid red curves: $A_X(\omega)$; short-dashed blue curves: 
$A_\Gamma(\omega)$, long-dashed green curves: $A_M(\omega)$.
}\label{Am}\end{figure}

To illustrate the Mott transition in the limit of half-filling, we show in 
Fig.~\ref{Am} the spectral distributions obtained from the interacting cluster 
Green's function: $A_{m}(\omega)=-(1/\pi)\,{\rm Im}\,G^{cl}_{m}(\omega+i\gamma)$, 
where $\gamma=0.02$. These spectra can be evaluated without requiring analytic 
continuation from Matsubara to real frequencies. The total density of states
per spin is given by  $A(\omega)=[A_\Gamma(\omega)+A_M(\omega)+2A_X(\omega)]/4$.  
All cluster molecular orbitals contribute to the spectral weight near the Fermi 
level in the metallic phase for $\delta>0$, and to the upper and lower Hubbard 
bands in the Mott phase at $\delta=0$.

The evolution of these spectra as a function of doping supports the picture
conjectured long ago by Eskes {\it et al.}\cite {eskes} Upon hole doping,
spectral weight is transfered from the upper and lower Hubbard bands to states
just above $E_F$, in the lower part of the Mott gap. Since the spectral weight 
(per spin) of both Hubbard bands initially decreases like $(1-\delta)/2$, the 
states induced just above $E_F$ have weight $\delta$ 
(see also Ref.~\cite{phillips}).
This scenario is a remarkable consequence of strong dynamical correlations and 
differs fundamentally from the one in ordinary semiconductors, where states 
induced in the gap have weight $\delta/2$ per spin for total doping $\delta$.

\begin{figure}[t]  
\begin{center}
\includegraphics[width=4.5cm,height=6.5cm,angle=-90]{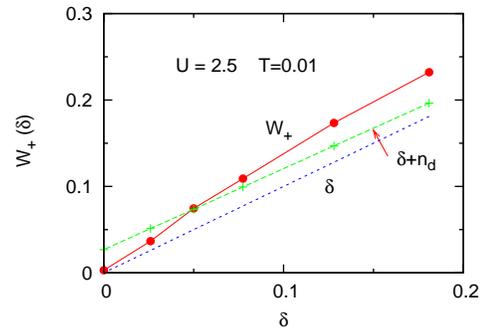}
\end{center}
\caption{(Color online)
Doping induced spectral weight $W_+(\delta)$ above $E_F=0$ up to about 
$\omega=0.9$, i.e., in the lower part of the main gap. 
A small constant weight is subtracted to account for the artificial 
broadening of the spectral peaks. This constant is chosen so that $W_+=0$
at $\mu=0.7$, $\delta=0$. The short-dashed blue line defines $\delta$ and 
the long-dashed green line  $\delta+n_d$, where $n_d$ is the average
double occupancy shown in Fig.~3.     
}\label{weight}\end{figure}

Our ED/DMFT cluster calculations are in excellent agreement with this picture,
as illustrated in Fig.~\ref{weight} which shows the integrated spectral weight 
per spin induced just above $E_F$. This weight is denoted here as $W_+(\delta)$. 
The initial slope of $W_+$ is seen to be well represented by $\delta$, 
confirming the scenario discussed above.
At finite doping $W_+(\delta)$ becomes even larger than $\delta + n_d$,
where $n_d$ is the double-occupancy shown in Fig.~\ref{nvsmu}. These results differ
from those for $t'=0$ and $T=0$ obtained by Sakai {\it et al.}\cite{sakai},
who found $W_+ (\delta) \approx \delta + n_d$ up to  about 14~\% hole doping. 

Upon closer inspection, the spectral distributions shown in Fig.~\ref{Am} at 
finite doping reveal a pseudogap close to $E_F$ which will be discussed in 
more detail in the following subsections. As shown below, this pseudogap is
intimately related to the non-Fermi-liquid properties which are evident in 
the $X$ component of the self-energy.

\subsection{Non-Fermi-Liquid Properties}

\begin{figure}[t]  
\begin{center}
\includegraphics[width=4.5cm,height=6.5cm,angle=-90]{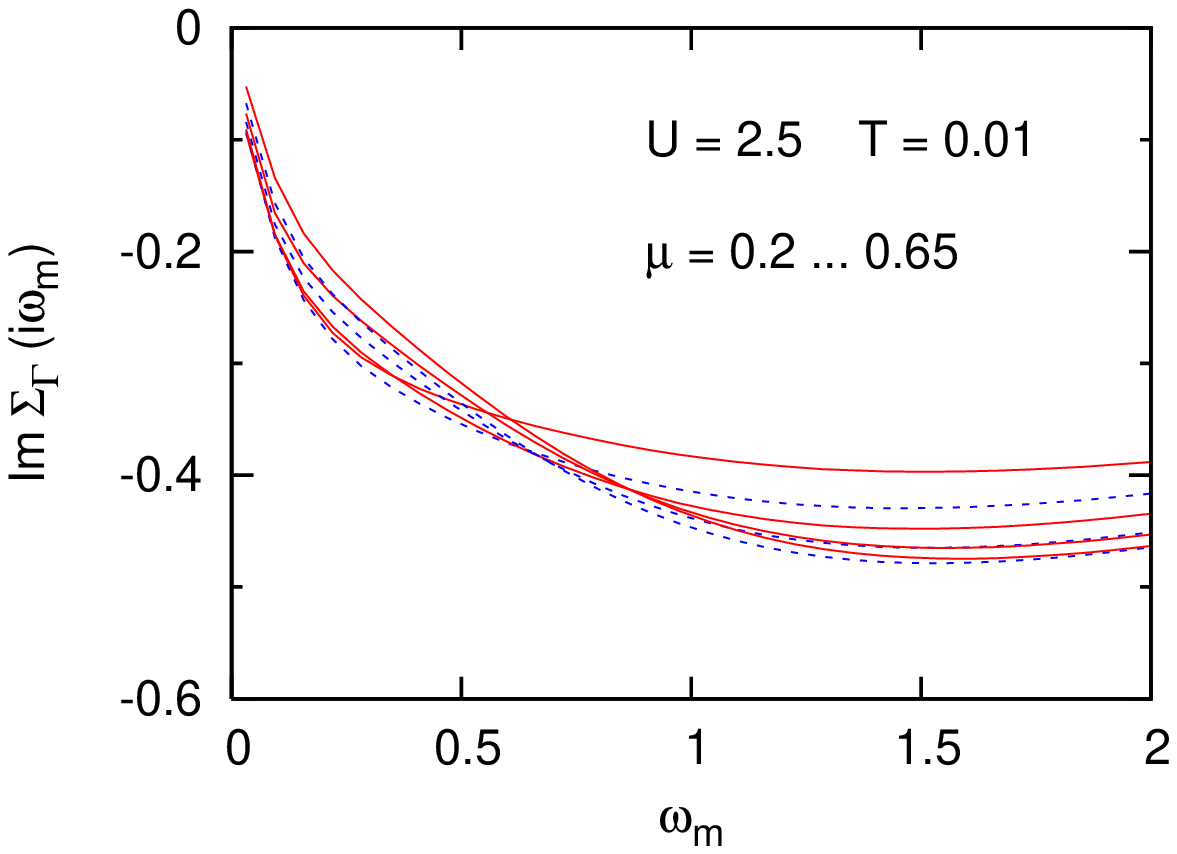}
\includegraphics[width=4.5cm,height=6.5cm,angle=-90]{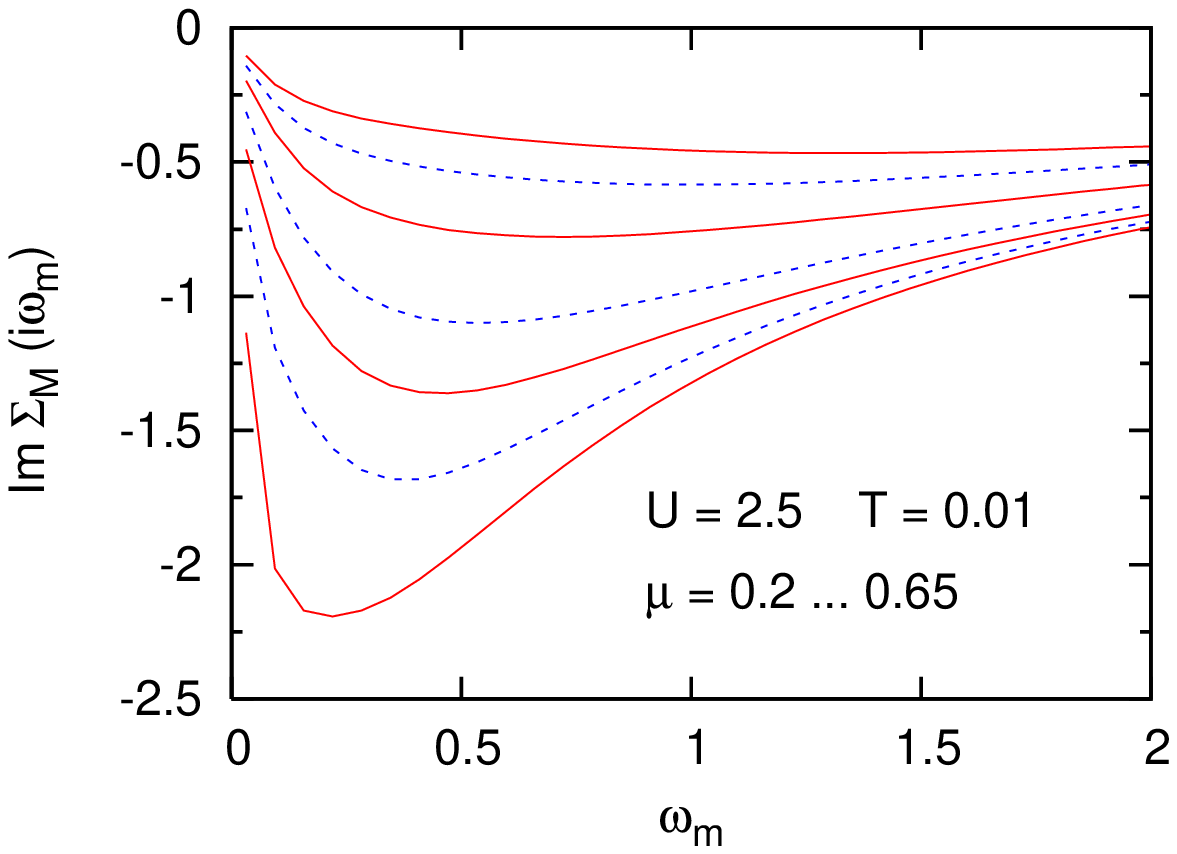}
\includegraphics[width=4.5cm,height=6.5cm,angle=-90]{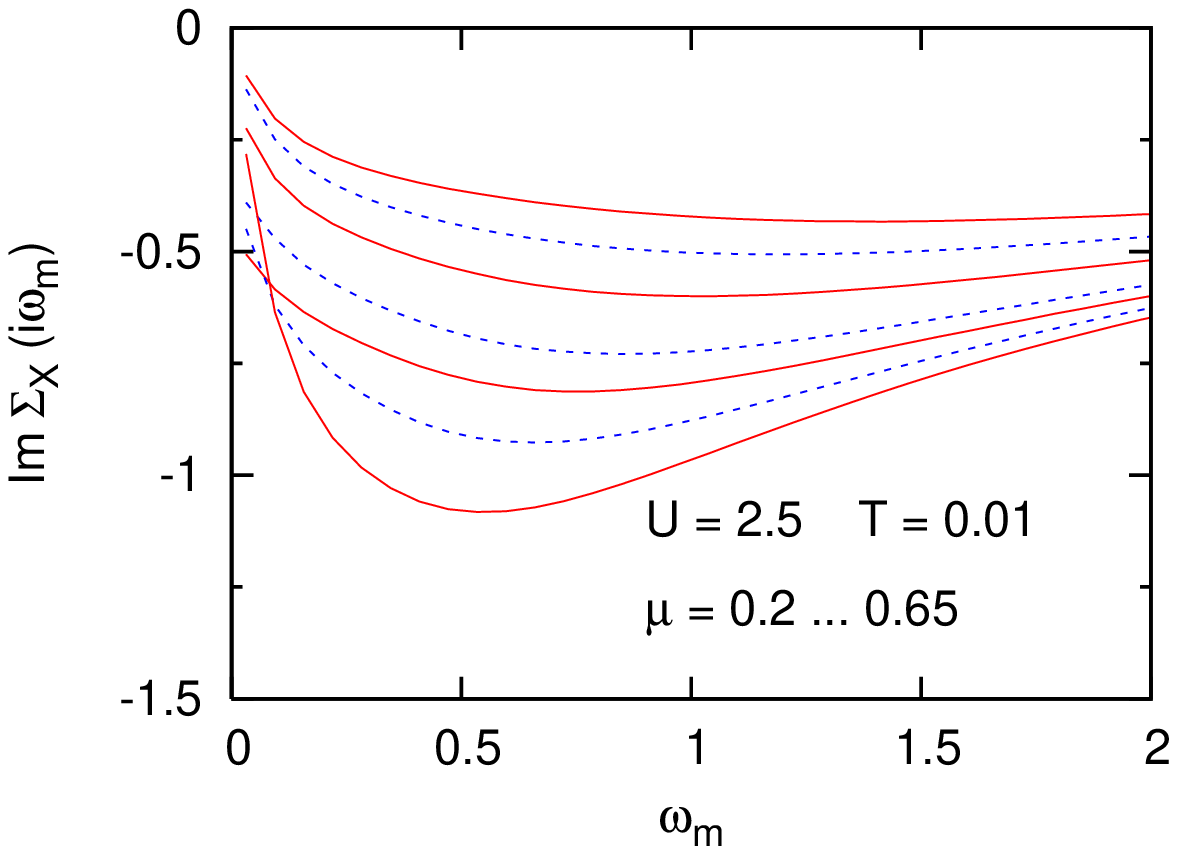}
\end{center}
\caption{(Color online)
Imaginary part of self-energy molecular orbital components 
$\Sigma_m(i\omega_n)$ as functions of Matsubara frequency for various
chemical potentials: 
$\mu=    0.2 ,\ 0.3 ,\ 0.4 ,\ 0.5 ,\ 0.55,\ 0.6 ,\ 0.65$ (from top to bottom) 
corresponding to $\delta= 0.24,\ 0.18,\ 0.12,\ 0.08,\ 0.05,\ 0.03,\ 0.01$, 
respectively; $U=2.5$, $T=0.01$.  
}\label{sigma}\end{figure}

\begin{figure}[t]  
\begin{center}
\includegraphics[width=4.5cm,height=6.5cm,angle=-90]{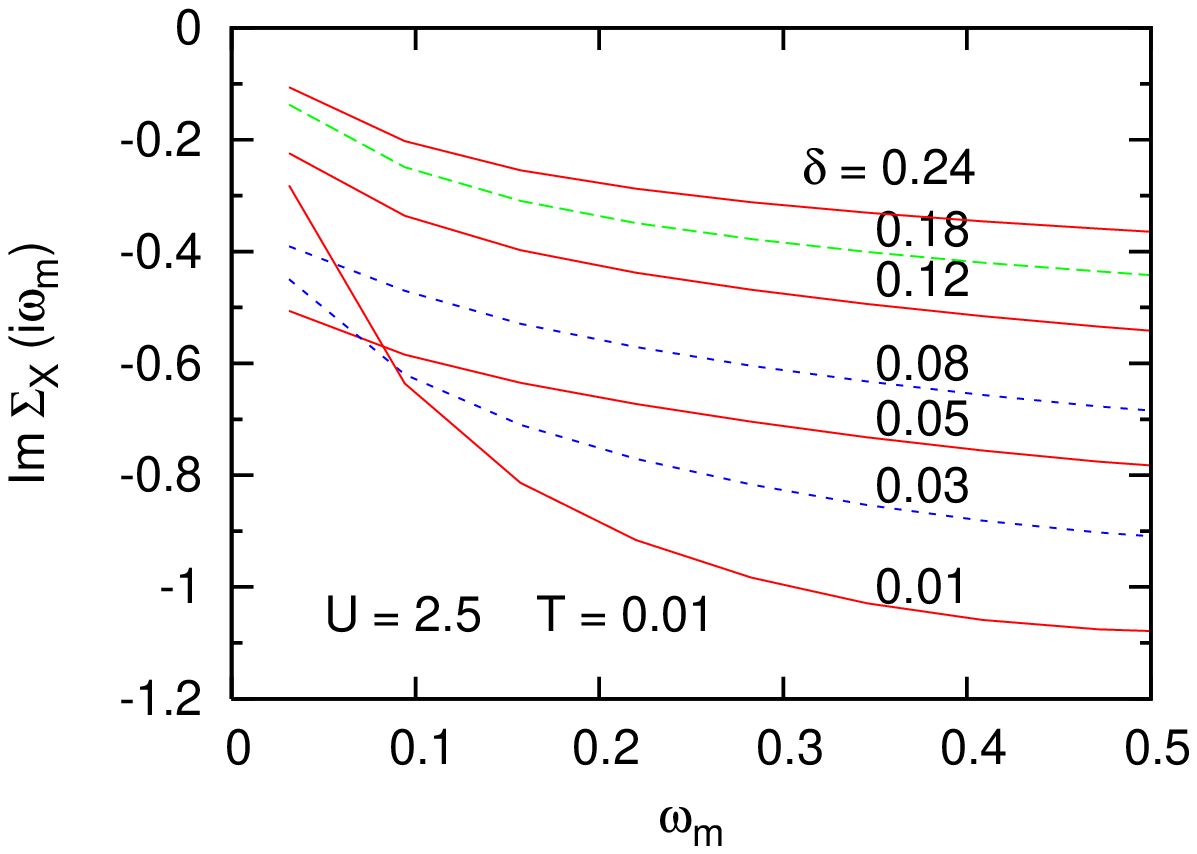}
\includegraphics[width=4.5cm,height=6.5cm,angle=-90]{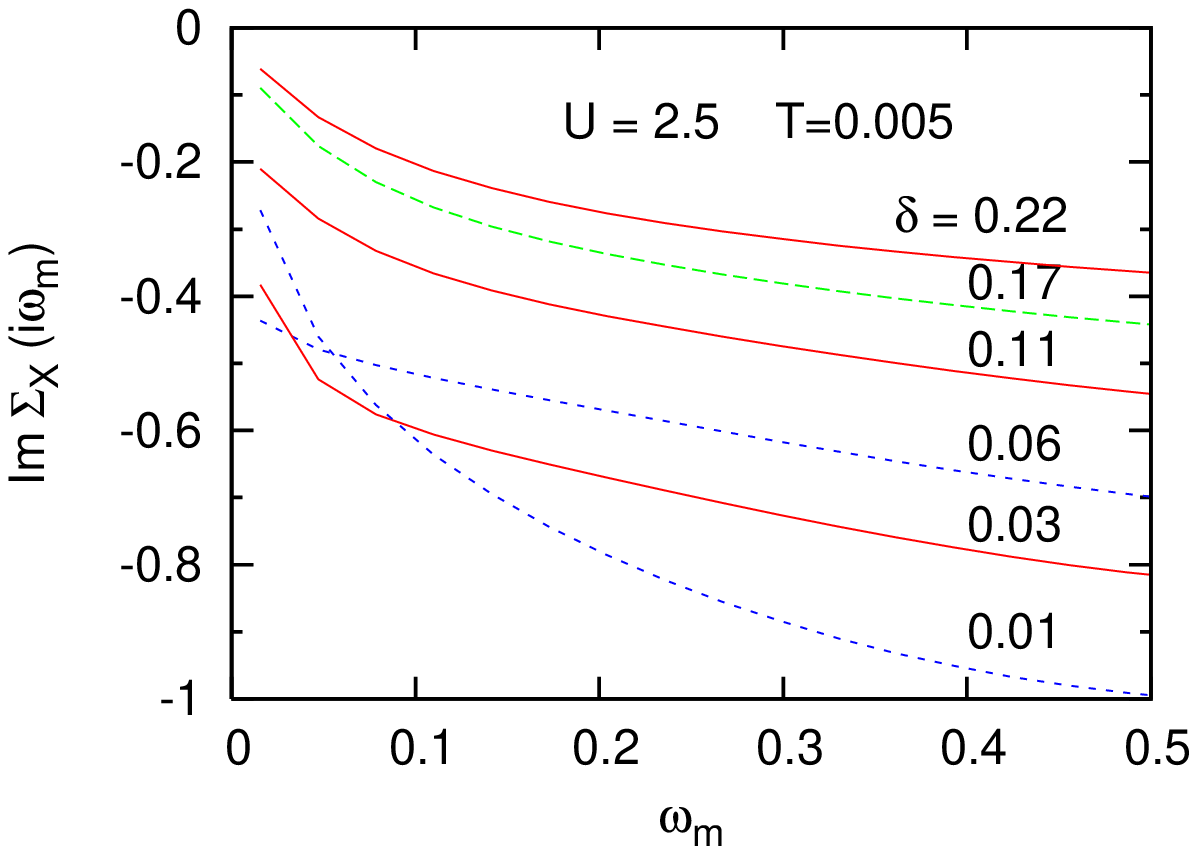}
\end{center}
\caption{(Color online)
Im\,$\Sigma_X(i\omega_n)$ for various hole doping concentrations on expanded 
scale for $T=0.01$ (upper panel) and $T=0.005$ (lower panel). 
The long-dashed green curves near $\delta\approx 18$~\% denote the approximate 
onset of non-Fermi-liquid behavior which grows until $\delta$ decreases to 
about 5~\%. 
}\label{sigmaT}\end{figure}

We now discuss the low-frequency variation of the cluster self-energy
which is strikingly different for the different cluster molecular orbitals.
Fig.~\ref{sigma} shows the imaginary parts of $\Sigma_m(i\omega_n)$, Eq.~(\ref{S}),
for chemical potentials $\mu$ in the range from $0$ to $24$~\% hole doping.
The $\Gamma$ orbital, approximately representative of the center of the 
Brillouin zone, exhibits the weakest self-energy. It is nearly independent of 
doping and Fermi-liquid-like, with only a moderate effective mass enhancement. 
$\Sigma_M$ changes from Fermi-liquid behavior at large doping to nearly 
insulating behavior $\sim 1/i\omega_n$ close to the Mott transition. 
At small finite doping, it reveals strong effective mass enhancement. 
Both Im\,$\Sigma_\Gamma$ and Im\,$\Sigma_M$ extrapolate to very small
finite values in the limit $\omega_n\rightarrow 0$, except near the Mott 
transition. In striking contrast to these orbitals, Im\,$\Sigma_X(i\omega_n)$ 
exhibits a finite onset in the low-frequency limit once the doping 
is smaller than about $20$~\% (see expanded scale in Fig.~\ref{sigmaT}). 
The onset is largest at about $\mu=0.55$, corresponding to $\delta=5$~\%. 
At smaller doping (larger $\mu$), i.e., very close to the Mott transition,
it diminishes again.

In addition to the low-frequency onset of Im\,$\Sigma_X(i\omega_n)$ which
gives rise to reduced quasiparticle lifetime, the non-Fermi-liquid behavior
also leads to a characteristic flattening of Im\,$\Sigma_X(i\omega_n)$, 
which induces a sharp resonance in Im\,$\Sigma_X(\omega)$ at small positive 
frequencies. As will be discussed in the next subsection, it is this 
resonance that is responsible for the pseudogap in the density of states.

Similar results are obtained at lower temperature, $T=0.005$, as shown 
in the lower panel of Fig.~\ref{sigmaT}. Again, the largest deviation 
from Fermi-liquid behavior is found for Im\,$\Sigma_X(i\omega_n)$ at 
about $6$~\% doping. The onset of non-Fermi-liquid properties occurs 
at slightly smaller doping than for $T=0.01$. The results shown in 
Figs.~\ref{sigma} and \ref{sigmaT} are consistent with the $T=0$ \
ED/CDMFT calculations by Civelli {\it et al.}\cite{civelli}           

\begin{figure}[t]  
\begin{center}
\includegraphics[width=4.5cm,height=6.5cm,angle=-90]{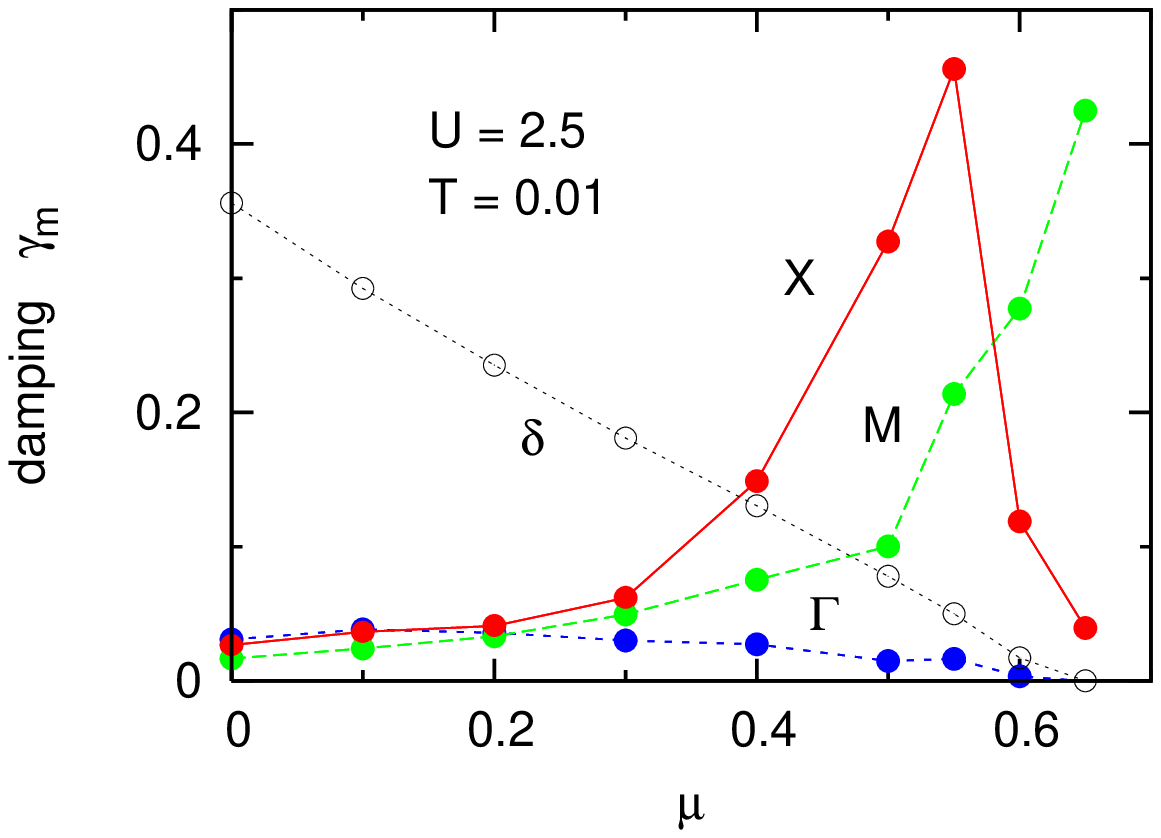}
\includegraphics[width=4.5cm,height=6.5cm,angle=-90]{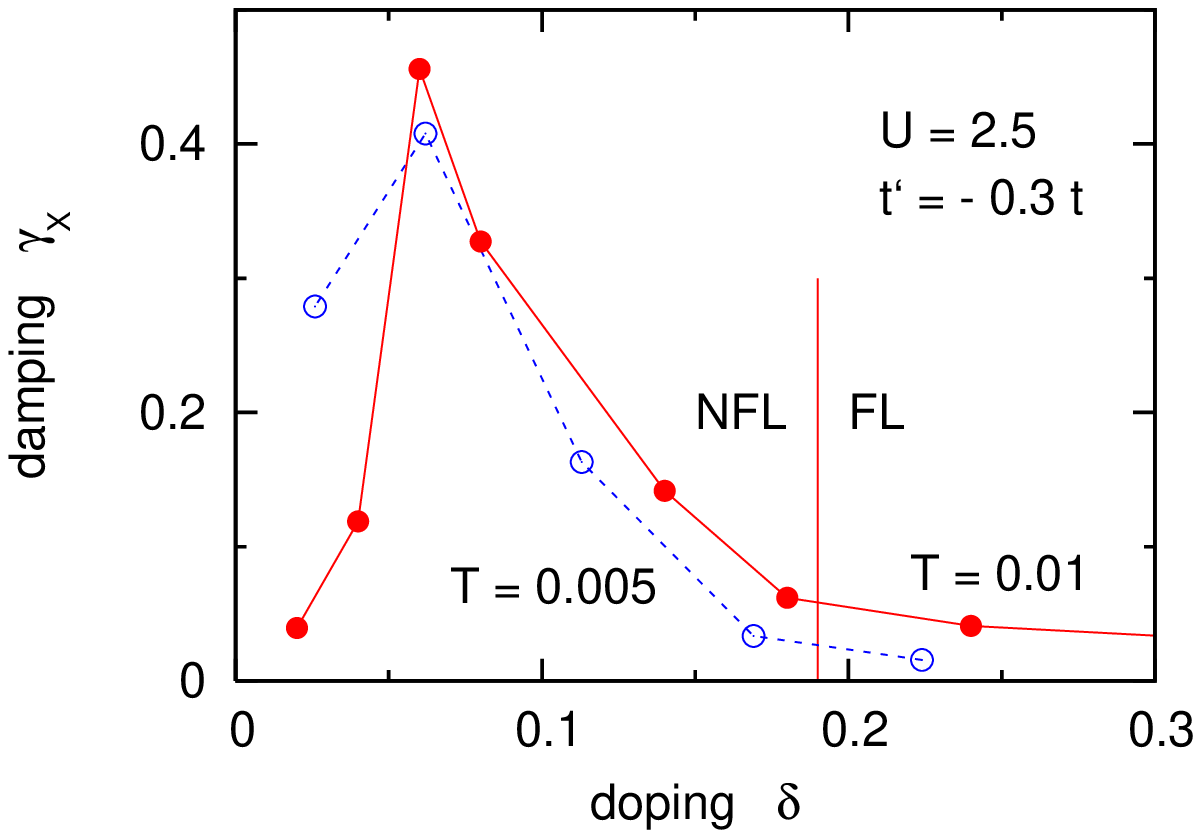}
\end{center}
\caption{(Color online) Upper panel:
Low-frequency damping rates 
$\gamma_m =-{\rm Im}\,\Sigma_m(i\omega_n\rightarrow0)$ 
as functions of chemical potential. Solid red curves: $X$ orbital, 
short-dashed blue curves: $\Gamma$ orbital, long-dashed green curves: 
$M$ orbital. The dotted curve denotes the doping $\delta$ (same scale 
as $\gamma_m$); $U=2.5$, $T=0.01$.  
For $\mu>0.2\ldots0.3$ or $\delta<0.18\ldots0.20$, $\gamma_X$ increases 
strongly, indicating the onset of non-Fermi-liquid behavior.
Lower panel: comparison  of $\gamma_X$ as a function of doping for 
$T=0.01$ (solid red circles) and $T=0.005$ (empty blue circles). 
The vertical bar denotes the approximate location
of the transition from Fermi-liquid to non-Fermi-liquid behavior.  
}\label{damping}\end{figure}

To illustrate this non-Fermi-liquid behavior of $\Sigma_X$ in more detail,
we compare in the upper panel of Fig.~\ref{damping} the low-frequency limits 
$\gamma_m \equiv -{\rm Im}\,\Sigma_m(i\omega_n \rightarrow 0)$  as functions 
of chemical potential. These values were found to be nearly the same for a 
linear extrapolation from the first two Matsubara points and for a 
quadratic fit using the first three points. At $\mu>0.3$ or $\delta<0.18$, 
$\gamma_X$ increases strongly, indicating the onset of a non-Fermi-liquid 
phase. The lower panel shows the variation of $\gamma_X$ with doping for
$T=0.01$ and $T=0.005$. At lower $T$, the onset of non-Fermi-liquid behavior 
is seen to be slightly sharper and to shift to slightly lower $\delta$.    

At finite temperature, a sharp transition between Fermi-liquid and
non-Fermi-liquid phases is not to be expected. According to the detailed 
temperature variation of the self-energy of the two-dimensional Hubbard
model studied recently by Vidhyadhiraja {\it et al.}\cite{vidhya} within 
QMC/DCA for $4\times4$ clusters ($U=1.5$, $t'=0$), a quantum critical point 
exhibiting marginal Fermi-liquid behavior was found at $\delta_c\approx 15$~\% 
doping, with Fermi-liquid behavior at larger $\delta$ and a pseudogap phase 
at $\delta < \delta_c$. At $T=0.01$, the $T$ / $\delta$ phase diagram 
indicates a crossover region of about $\delta=0.15\pm 0.02$ between these 
phases. Assuming a crossover region of similar width, i.e., 
$\Delta\delta\approx0.04$, the results shown in Fig.~\ref{damping} are 
consistent with those of Ref.\cite{vidhya}. Thus, for $U=2.5$, $t'=-0.3 t$,
the Fermi-liquid and non-Fermi-liquid phases seem to be separated by a 
quantum critical point at $\delta_c\approx 18\ldots 20$~\%, with marginal 
Fermi-liquid behavior for $T\ge0$.

\begin{figure}[t]  
\begin{center}
\includegraphics[width=4.5cm,height=6.5cm,angle=-90]{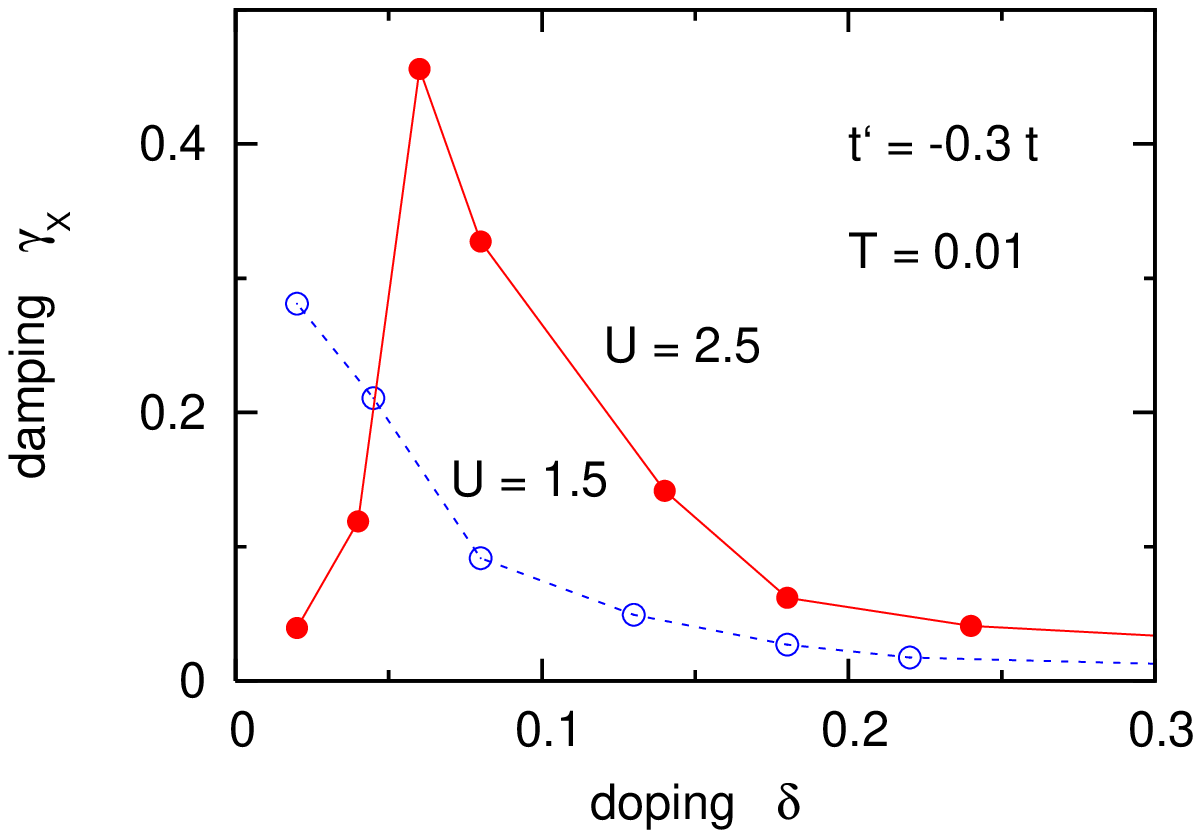}
\includegraphics[width=4.5cm,height=6.5cm,angle=-90]{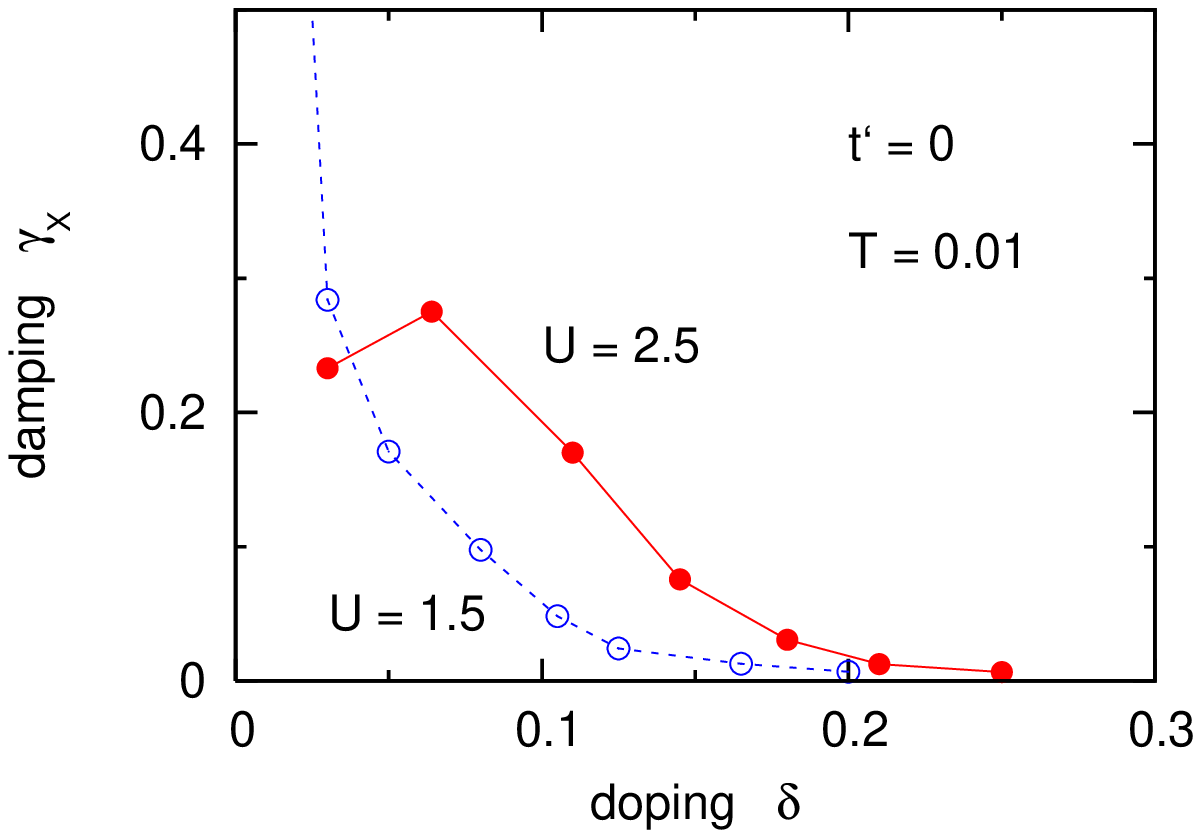}
\end{center}
\caption{(Color online) Low-frequency damping rate of $X$ orbital, 
$\gamma_X = -{\rm Im}\,\Sigma_X (i\omega_n\rightarrow0)$ as a function of 
doping for various Coumb energies and next-nearest-neighbor hopping energies; 
$T=0.01$.
Upper panel: $t'=-0.3t$, lower panel: $t'=0$. 
Solid red circles: $U=2.5$, empty blue circles: $U=1.5$. 
}\label{Gnud}\end{figure}
 
Figure~\ref{Gnud} compares the low-frequency damping rate of the $X$ orbital 
as a function of doping for $U=2.5$ and $U=1.5$. The upper panel shows the 
results for $t'=-0.3t$, the lower panel for $t'=0$. The case $U=1.5$, $t'=0$ 
suggests a critical doping $\delta_c\approx 0.15 \pm 0.02$, in
agreement with the results of Ref.~\cite{vidhya}.  As is to be expected,
at smaller $U$  $\delta_c$  is smaller than at large $U$, since the Fermi 
liquid properties are stabilized. A similar trend occurs as $t'$ is shifted 
from $t'=-0.3t$ \,to \,$t'=0$. Nevertheless, despite the large variations in 
$U$ and $t'$, the critical doping separating the Fermi-liquid and 
non-Fermi-liquid phases is remarkably stable and occurs in the range 
$\delta_c\approx 0.15\ldots 0.20$, i.e., close to the optimal doping 
concentrations found in many high-$T_c$ cuprates.  

\begin{figure}[t]  
\begin{center}
\includegraphics[width=4.5cm,height=6.5cm,angle=-90]{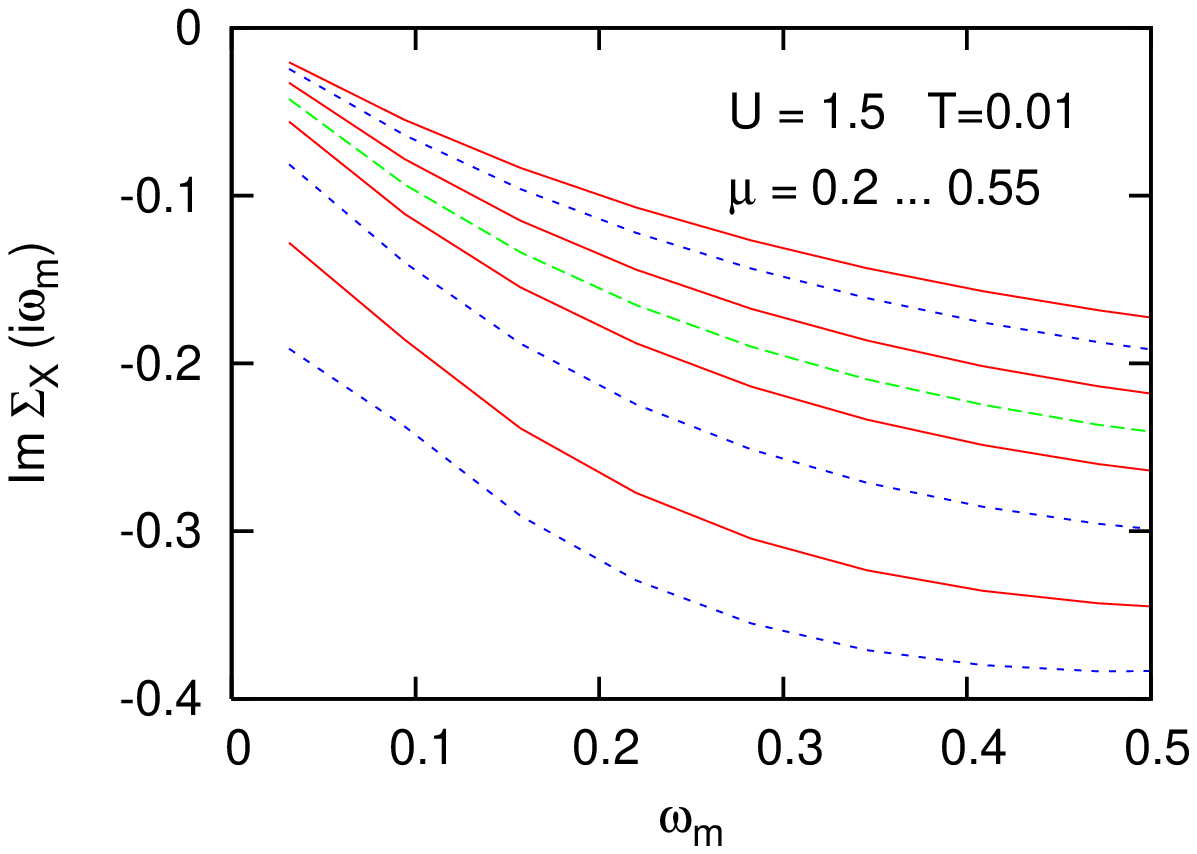}
\includegraphics[width=4.5cm,height=6.5cm,angle=-90]{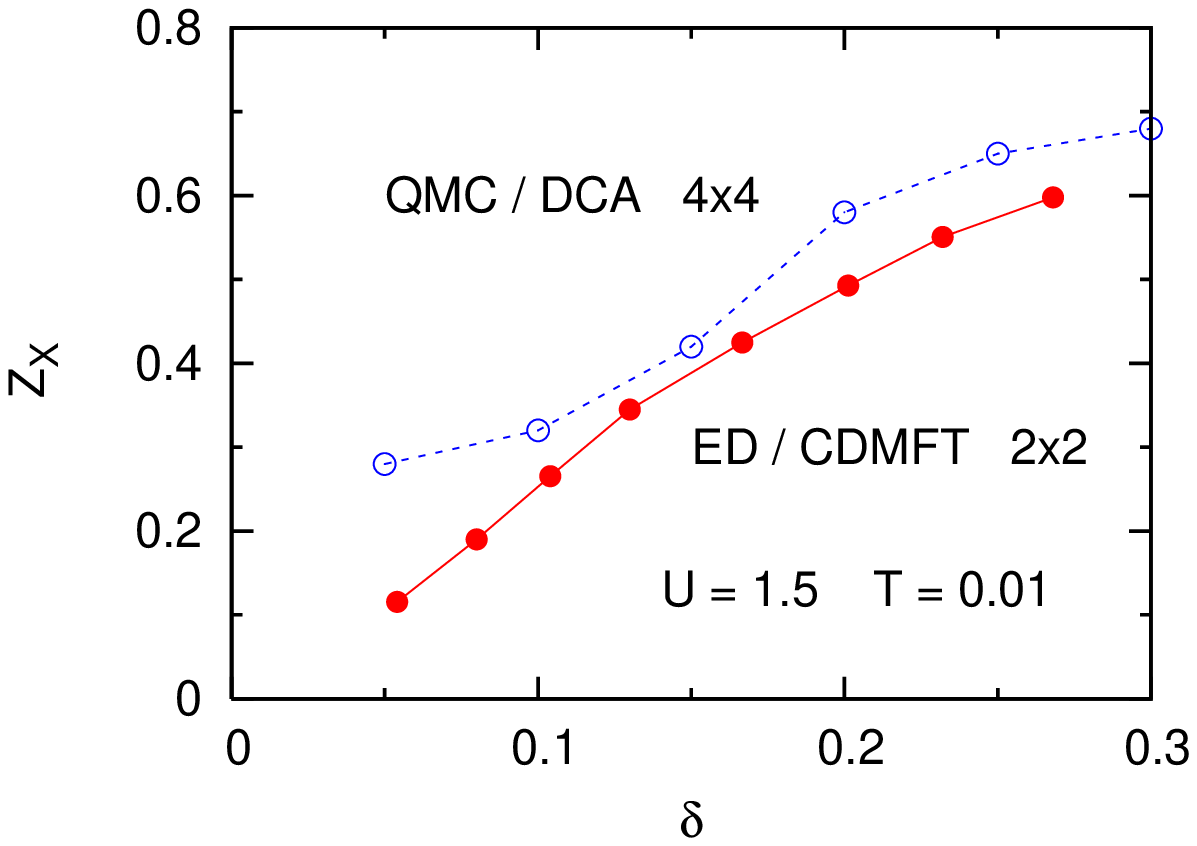}
\end{center}
\caption{(Color online) Upper panel: imaginary part of $\Sigma_X (i\omega_n)$ 
as a function Matsubara frequency for $\mu = 0.2 \ldots 0.55$ in steps of $0.05$
(from top), for  $U=1.5$, $t'=0$, $T=0.01$. $\mu =0.3$ (long-dashed green curve) 
corresponds to $\delta=0.18$ and approximately marks the transition from 
Fermi-liquid to non-Fermi-liquid behavior. Lower panel: Solid red circles: 
$Z_X = 1/(1-{\rm Im}\Sigma_X(i\omega_0)/\omega_0)$ derived from ED/CDMFT 
results in upper panel; empty blue circles: analogous QMC/DCA results from 
Fig.~1 of Ref.~\cite{vidhya} for $T=0.014$.   
}\label{jarrell}\end{figure}
         
For a more detailed comparison with the results of Ref.~\cite{vidhya}, we show
in Fig.~\ref{jarrell} the variation of Im\,$\Sigma_X(i\omega_n)$ with chemical 
potential for $U=1.5$, $t'=0$ and $T=0.01$. These values of $\mu$ correspond to 
dopings in the range $\delta=0.27 \ldots 0.05$.  Although the overall magnitude of   
 Im\,$\Sigma_X$ is much smaller than in Fig.~\ref{sigmaT} for $U=2.5$, $t'=-0.3t$, 
there is again a clear separation between doping larger than $\delta_c\approx 0.17$
exhibiting Fermi-liquid behavior, and smaller doping with characteristic          
non-Fermi-liquid features. The lower panel shows the comparison of the approximate
quasiparticle weight, $Z_X = 1/[1-{\rm Im}\Sigma_X(i\omega_0)/\omega_0]$, 
derived from the ED/DMFT results in the upper panel, with the corresponding 
QMC/DCA values taken from Fig.~1 of Ref.~\cite{vidhya}. For 
$\delta>0.15$ the agreement is very good. (Note that for $Z_X>0.5$, 
$-{\rm Im}\,\Sigma_X(i\omega_0)$ is less than $\omega_0=0.031$.)
At smaller doping, the difference becomes larger, presumably because of the 
finer momentum resolution achieved for the $4\times4$ cluster in Ref.~\cite{vidhya}.

To analyze the difference between CDMFT and DCA for $2\times2$ clusters, with 
identical system parameters, we compare in Fig.~\ref{dca} the low-frequency 
damping rate of $\Sigma_X (i\omega_n)$ as a function of doping. 
Evidently, the different relations between self-energy components $\Sigma_m$ 
and lattice Green's function $G_m$ in these two schemes give rise to changes 
on a quantitative level. Nevertheless, both approaches predict a transition 
from a Fermi-liquid phase at hole doping larger than about 20~\% to a 
non-Fermi-liquid phase at small doping. Surprisingly, the transition is less 
sharp in DCA than in CDMFT. The reason for this difference might be that, in
contrast to CDMFT, the momentum patches of the Brillouin zone are not coupled 
in the evaluation of the DCA lattice Green's function (see Eq.~(\ref{Gdca})). 
It might therefore be necessary in DCA to treat larger clusters 
(such as $8$ sites\cite{werner} or $16$ sites\cite{vidhya}) 
in order to obtain a sharper Fermi-liquid to non-Fermi-liquid transition. 
A slower convergence with cluster size in DCA is also found for the critical 
Coulomb energy at half filling.\cite{gull,werner}

\subsection{Pseudogap}

The non-Fermi-liquid properties of $\Sigma_X(i\omega_0)$ manifest themselves
not only in the enhanced low-frequency damping rate discussed above, but also
in the flattening of Im\,$\Sigma_X(i\omega_n)$ which can be identified as the 
origin of the pseudogap in the density of states. Narrow gaps near $E_F$ below 
the critical doping are already evident in the cluster spectra shown in 
Fig.~\ref{Am}. Fig.~\ref{A} shows these spectra on an expanded scale for 
$\delta=\delta_c\approx0.18$ and $\delta=0.03$. While near critical doping 
the density of states is Fermi-liquid-like, with a sharp peak at $E_F$, 
smaller hole doping leads to a very asymmetric density of states, with 
a pseudogap of magnitude $\Delta \approx 4 t^2/U=0.1$ right above $E_F$.   
The molecular orbital analysis of these spectra reveals 
that this pseudogap is associated entirely with the $A_X(\omega)$ 
contribution, i.e., with scattering processes involving momenta close to
$(\pi,0)$ and $(0,\pi)$. With decreasing doping, the peak at $E_F$ seen for
$\delta\approx\delta_c$ shifts downwards, so that the Fermi level
gradually moves into the pseudogap. At the same time, the pseudogap becomes
wider and the spectral weight above $E_F$ is reduced until the transition 
to the Mott phase occurs at half filling. (The peak at $\omega\approx 0.25$
for $\delta=0.18$ is due to the discreteness of the cluster spectra and 
is not related to the pseudogap. The actual pseudogap at this large doping 
is vanishingly small; see analysis of self-energy below).

\begin{figure}[t]  
\begin{center}
\includegraphics[width=4.5cm,height=6.5cm,angle=-90]{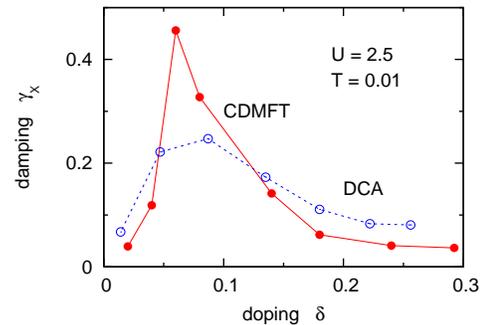}
\end{center}
\caption{(Color online) 
Comparison of low-frequency damping rate of $X$ orbital, 
$\gamma_X = -{\rm Im}\,\Sigma_X (i\omega_n\rightarrow0)$ as a function of 
doping for CMDFT (solid red circles) and DCA (empty blue circles);
$U=2.5$, $T=0.01$.
}\label{dca}\end{figure}

Note that the peak at $E_F$ for $\delta=\delta_c$ 
is also compatible with marginal Fermi-liquid behavior.
Finite-size effects, however, do not permit a clear distinction between 
Fermi-liquid properties below the first Matsubara frequency $\omega_0$ and 
genuine marginal Fermi-liquid behavior at $\delta_c$.\cite{vidhya}      
              
The lower panel of Fig.~\ref{A} shows the corresponding spectra
derived from the lattice Green's function components $G_m(i\omega_n)$,  
Eq.~(\ref{Gm}), via extrapolation to real $\omega$. Thus, $A(\omega)=
-\frac{1}{\pi}{\rm Im}\,[G_\Gamma(\omega)+G_M(\omega)+2G_X(\omega)]/4$.  
We use here the routine {\it ratint}.\cite{ratint} 
Nearly identical spectra are obtained via Pad\'e extrapolation.
About $400\ldots600$ Matsubara points are taken into account for the energy 
window $-1\le\omega\le 1$, and the same broadening is assumed ($\gamma=0.02$) 
as in the cluster spectra shown in the upper panel. 
As a result of the accurate self-energies and Green's functions along the
Matsubara axis, the extrapolation to low real $\omega$ is highly reliable.
The lattice spectra confirm the trend observed in the cluster spectra: 
At $\mu=0.3$, $\delta=0.18$, the density of states has a peak very close 
to the Fermi level, while for  $\mu=0.6$, $\delta=0.03$, $E_F$ lies
in a pseudogap of about the same width as in the cluster data. The lattice
spectra $A(\omega)$ can also be calculated by first extrapolating the self-energy
components $\Sigma_m(i\omega_n)$ to real frequencies and then using Eq.~(\ref{Gm})
at real $\omega$. The results are fully consistent with the spectra derived
via extrapolation of $G_m(i\omega_n)$.

\begin{figure}[t]  
\begin{center}
\includegraphics[width=4.5cm,height=6.5cm,angle=-90]{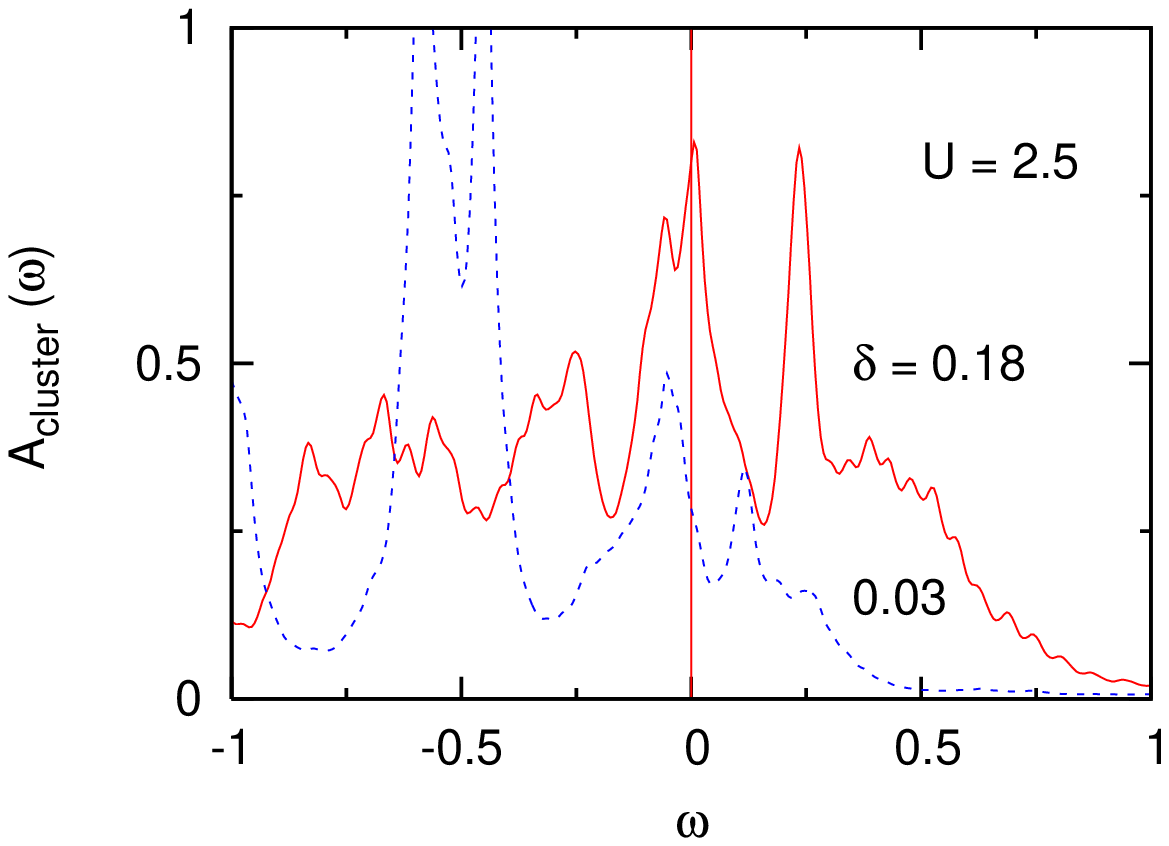}
\includegraphics[width=4.5cm,height=6.5cm,angle=-90]{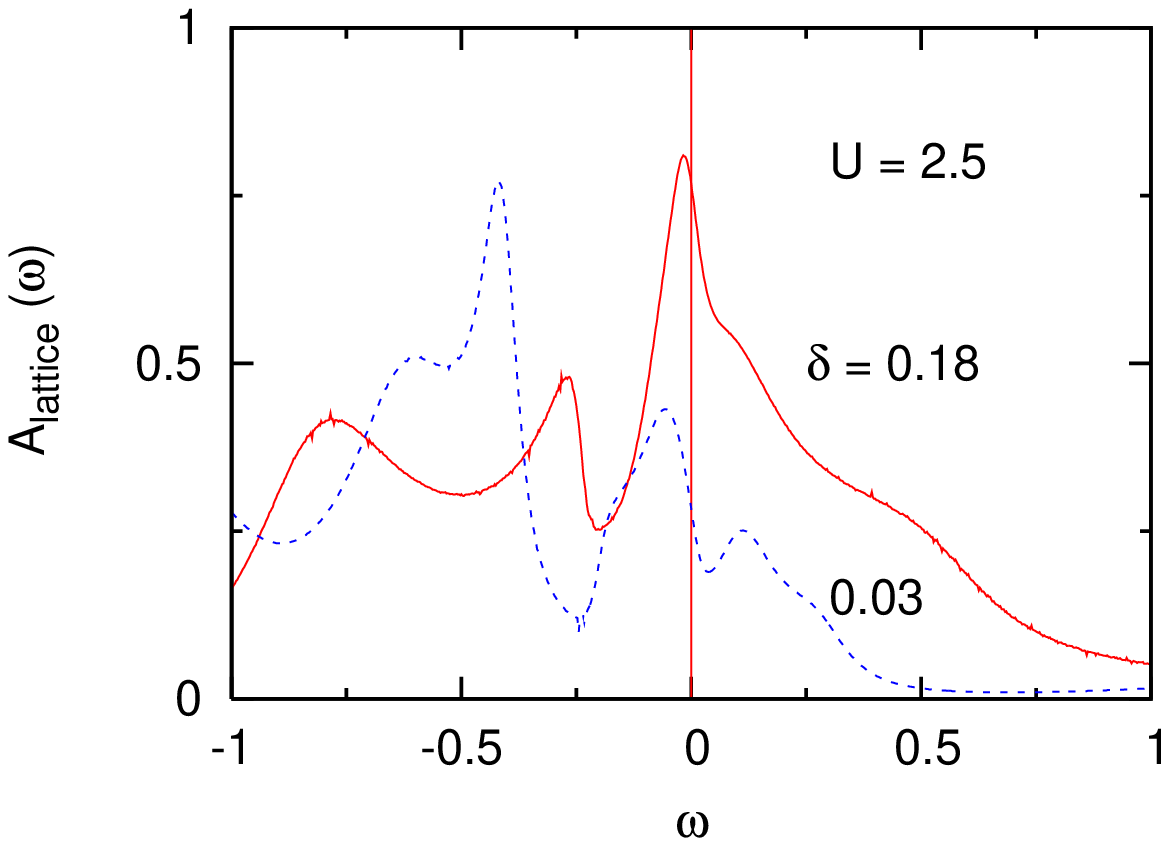}
\end{center}
\caption{(Color online)
Upper panel: Cluster spectral distributions 
$A(\omega)=\frac{1}{4}[A_\Gamma(\omega)+A_M(\omega)+2A_X(\omega)]$ for $U=2.5$, 
$t'=-0.3t$, $T=0.01$. 
Solid red curve: $\mu=0.3$ (hole doping $\delta=0.18$) with peak at $E_F=0$;  
dashed blue curve: $\mu=0.6$ ($\delta=0.03$) exhibiting a pseudogap above $E_F$;
lower panel: analogous lattice spectra obtained via extrapolation of 
Green's function components, Eq.~(\ref{Gm}).  
}\label{A}\end{figure}

The pseudogap seen in Fig.~\ref{A} for $\delta=0.03$ is reminiscent of the 
pseudogap obtained in the two-band model within local DMFT above the first 
Mott transition.\cite{prb04} 
Once the electrons in the narrow subband are Mott localized, an effective
two-fluid system is realized in which the Coulomb interaction with the 
remaining conduction electrons generates deviations from Fermi-liquid behavior, 
in particular, the finite lifetime associated with the low-frequency limit 
of ${\rm Im}\,\Sigma(i\omega_n)$, and the characteristic flattening of this 
function which gives rise to a pseudogap at real $\omega$.\cite{al+costi} 
This two-band model exhibits a quantum critical point when the pseudogap 
turns into the Mott gap.\cite{costi}  
It would be interesting to inquire whether the present cluster picture 
of the single band model could be mapped onto this two-band model. 
The spatial degrees of freedom in the cluster would then play the role
of the inter-orbital fluctuations in the two-band model.  
Since at small hole doping a sizable number of electrons is Mott localized 
their spins act as scattering centers for the remaining electrons whose 
self-energy then exhibits deviations from Fermi-liquid behavior. 

\begin{figure}[t]  
\begin{center}
\includegraphics[width=4.5cm,height=6.5cm,angle=-90]{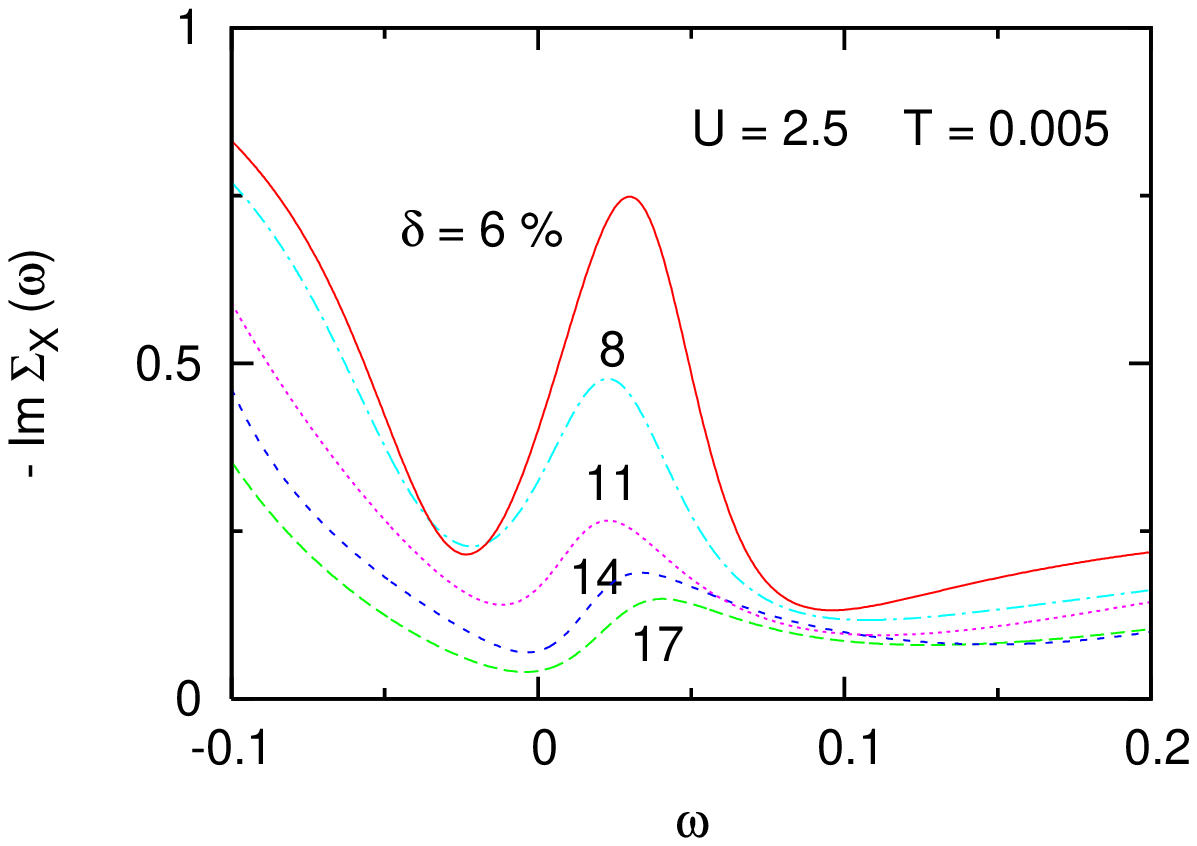}
\includegraphics[width=4.5cm,height=6.5cm,angle=-90]{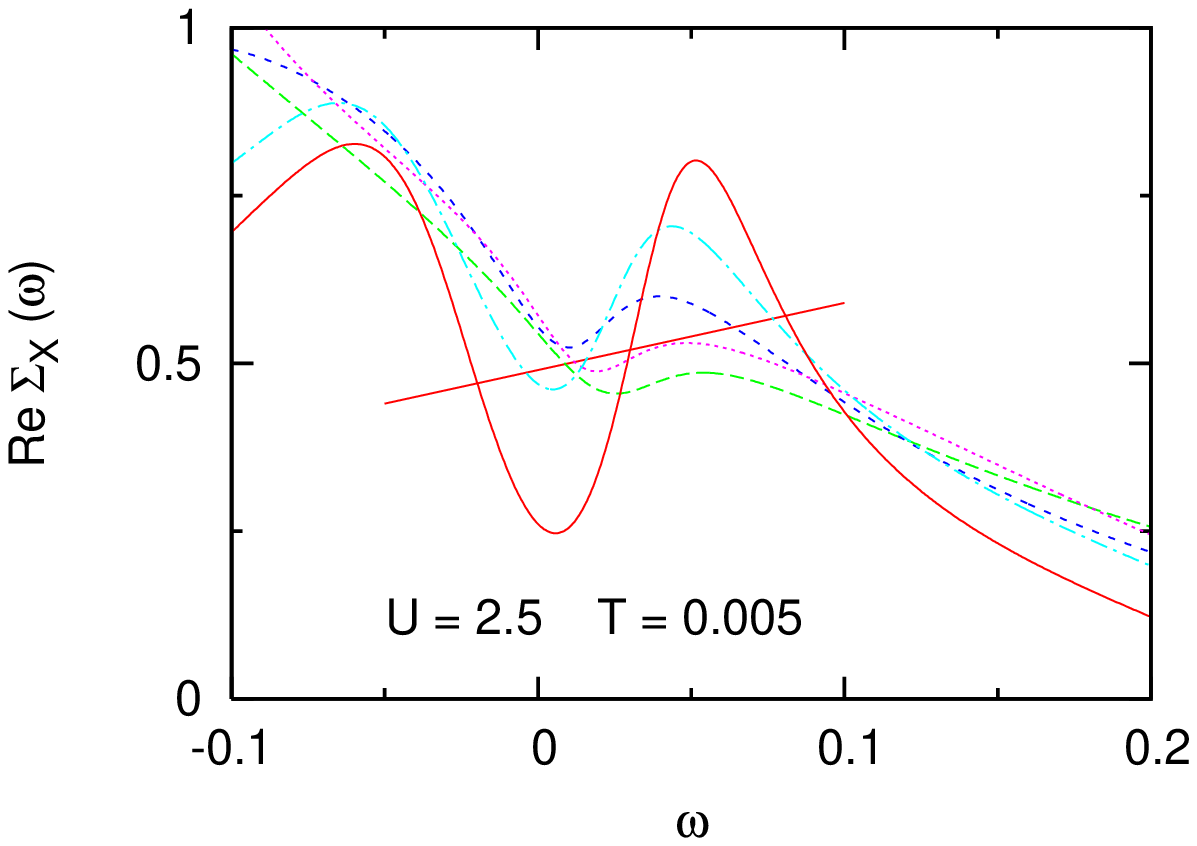}
\includegraphics[width=4.5cm,height=6.5cm,angle=-90]{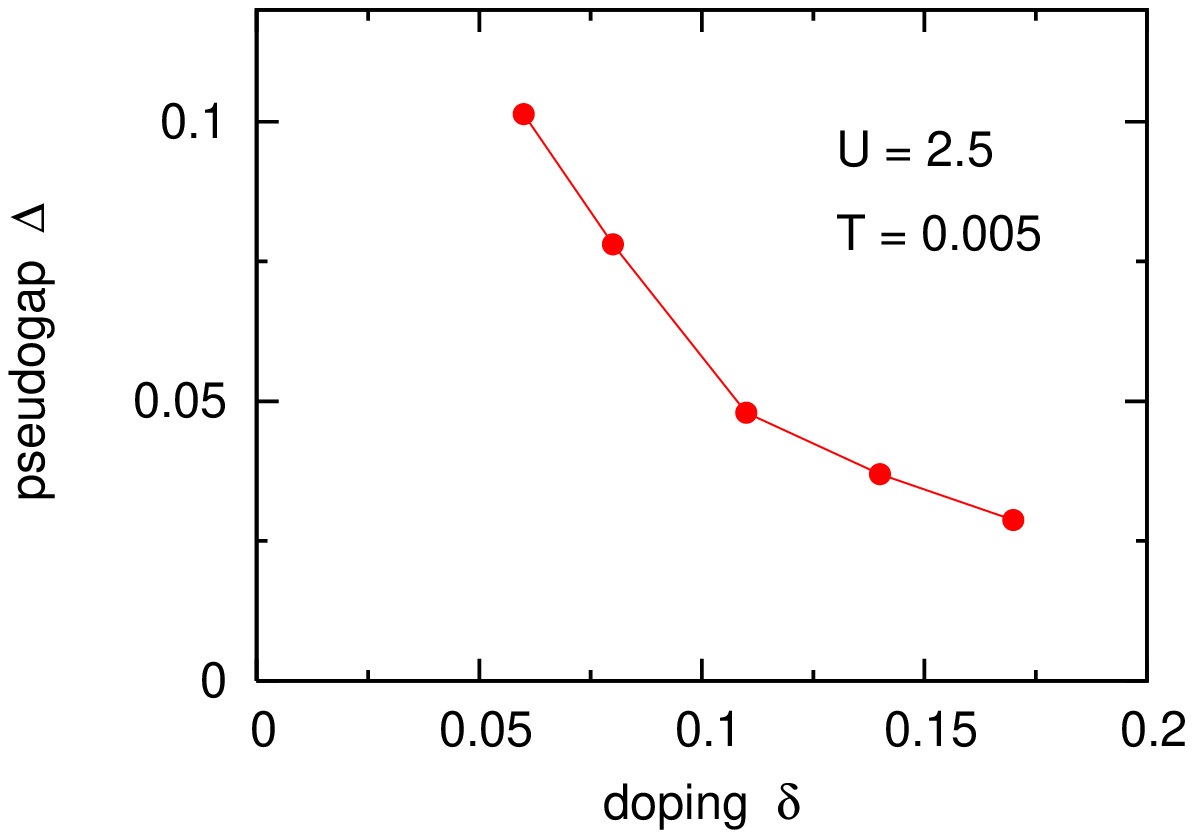}
\end{center}
\caption{(Color online)
Self-energy $\Sigma_X(\omega)$ obtained via extrapolation to real frequencies
for several hole doping concentrations (broadening $\gamma=0.005$).
Upper panels: $-$Im\,$\Sigma_X$ and Re\,$\Sigma_X$; $U=2.5$, $T=0.005$. 
The outer intersections of Re\,$\Sigma_X$ with the straight lines 
$\omega+\mu-\epsilon_k$ yield the approximate width of the pseudogap $\Delta$.
Lower panel: Pseudogap $\Delta$ as a function of hole doping 
derived from self-energies $\Sigma_X(\omega)$  shown in upper panels.
}\label{sr}\end{figure}

To illustrate the effect of non-Fermi-liquid behavior on the self-energy 
at real $\omega$, we show in Fig.~\ref{sr} the low-frequency variation 
of Im\,$\Sigma_X(\omega)$ obtained from the lower panel of Fig.~\ref{sigmaT} 
via extrapolation to real $\omega$. Typically, at these low frequencies we 
use the first $100\ldots400$ Matsubara points and evaluate Im\,$\Sigma_X$ at 
$\omega+i\gamma$, with $\gamma=0.005$. Although the details of the resulting 
spectra differ slightly, the important qualitative features near $E_F$ are 
very stable. Spectra derived via Pad\'e extrapolation are very similar.

As can be seen in Fig.~\ref{sr}, at large hole doping $-$Im\,$\Sigma_X(\omega)$ 
has a minimum at $E_F$ and varies quadratically at small $\omega$, as expected 
for a Fermi liquid. Damping in this range is very weak. Nevertheless, even for 
$\delta\ge 0.17$ a small peak in $-$Im\,$\Sigma_X(\omega)$ is found at about 
$\omega=0.05$ above $E_F$, indicating that electrons added to the system just 
above $E_F$ have a reduced lifetime. With decreasing doping, this feature grows 
into a strong resonance which eventually dominates the low-frequency properties. 
The minimum of $-$Im\,$\Sigma_X(\omega)$ is then shifted slightly below $E_F$ 
and a second minimum appears above $E_F$. Moreover, this resonance shows a 
dispersion as a function of doping. It first shifts downwards from 
$\omega\approx 0.05$ to $0.02$ and then disperses again upwards to about 
$\omega\approx0.05$. Re\,$\Sigma_X(\omega)$ is seen to exhibit a positive 
slope at the resonance which is consistent with Kramers-Kronig relations. 
This implies that spectral weight is removed from the resonance region where 
correlation induced damping is large. 

The evolution of the resonance in Im\,$\Sigma_X(\omega)$ with doping is one 
of the main results of this work and has to our knowledge not been discussed
previously. A weak resonance in Im\,$\Sigma_X$ at $5$~\%  doping was also 
found by Jarrell {\it et al.}\cite{jarrell2001} within QMC/DCA for $n_c=4$. 
The fact that this resonance is much stronger in the present results might 
be related to the faster convergence of CDMFT with cluster size (see  
discussion of Fig.~\ref{dca}).  A resonance in Im\,$\Sigma(\omega)$ is 
also obtained in the spectral weight transfer model proposed by 
Philipps {\it et al.}\cite{phillips} In this scheme, however, the 
resonance is located at $\omega=0$ independently of doping.    

The outer intersections of Re\,$\Sigma_X(\omega)$ with 
$\omega+\mu-\epsilon_k$ provide the approximate width of the pseudogap 
$\Delta$ in the spectral distribution. The central intersection does not 
yield any peak because of the short lifetime in this frequency range. 
The new minima of $-$Im\,$\Sigma_X(\omega)$ below and above the resonance 
are consistent with the spectral peaks just below and above $E_F$, as seen 
in the results for $\delta=0.03$ in  Fig.~\ref{A}. 
For increasing hole doping, the resonance 
of Im\,$\Sigma_X(\omega)$ becomes weaker so that for $\delta> 0.17$ there 
are no longer three intersections of $\omega+\mu-\epsilon_k$ with 
Re\,$\Sigma_X(\omega)$. The pseudogap then vanishes. At smaller doping, 
the peak in $-$Im\,$\Sigma_X(\omega)$ grows and the pseudogap gets wider. 
This trend, however, is superceded by the reduction of spectral weight 
above $E_F$ as the Mott transition at half-filling is approached.     

The lower panel of Fig.~\ref{sr} shows the approximate width of the 
pseudogap $\Delta$. 
We use here the outer intersections of Re\,$\Sigma_X(\omega)$ with the 
lines $\omega+\mu-\epsilon_k$ to define the magnitude of $\Delta$, where
$\epsilon_k$ is chosen so that $\omega+\mu-\epsilon_k$ passes through 
the inflection point in the region of the maximal positive slope of  
Re\,$\Sigma_X(\omega)$. Other values of $\epsilon_k$ yield similar 
values of $\Delta$. In the spectral distributions, this definition of
the pseudogap roughly corresponds to the peak-to-peak separation of 
spectral weight near the gap. Systematically smaller values of $\Delta$
are obtained, for instance, if the width of the gap half-way between the 
minimum of $A(\omega)$ and the neighboring maxima is chosen as definition. 
At $\delta> 0.17$, the definition used above no longer yields a pseudogap 
and the system turns into an ordinary Fermi liquid.

The doping dependent resonance in Im\,$\Sigma_X(\omega)$ and the concomitant 
opening of the pseudogap are consistent with recent angle-resolved 
photoemission (ARPES) data by Yang {\it et al.}~\cite{yang3} 
According to the upper panel of Fig.~\ref{sr}, for $\delta\ge 0.17$ the 
quasiparticle damping is symmetric for electron and hole states. 
Below this doping, the lifetime of electron states above $E_F$ is much 
shorter than that of hole states below $E_F$, giving rise to the opening 
of the pseudogap above $E_F$ and the striking particle-hole asymmetry 
observed in the data. Moreover, the results shown in Fig.~\ref{sr} are
specific to the $(\pi,0)$ component of the self-energy and are absent in 
$\Sigma_\Gamma(\omega)$ and $\Sigma_M(\omega)$. Thus, the particle-hole 
asymmetry and pseudogap above $E_F$ are momentum dependent features which 
are most pronounced in the antinodal region, but weak or absent along the
nodal $\Gamma M$ direction which also agrees with the experimental data.     
\cite{yang3} A more detailed discussion of the momentum variation of the
self-energy will be given in the final subsection.

Because of the finite temperature in the ED/CDMFT calculation, it is not
possible to identify spectral features at frequencies below the first
Matsubara point ($\omega_0=0.0314$ for $T=0.01$). 
Nevertheless, the doping variation of the pseudogap shown in Fig.~\ref{sr}
is found to be robust. In particular, it is clear that the pseudogap is
directly linked to the resonance in $-$Im\,$\Sigma_X(\omega)$ which, in 
turn, reflects the non-Fermi-liquid properties of the system. Since for
$\delta>\delta_c$ ordinary Fermi-liquid behavior is established, it is
evident that the pseudogap then vanishes.

The above scenario is consistent with the fact that for a hole-doped Mott 
insulator the addition of electrons pushes the system closer to the 
insulating phase. This implies that spectral weight just above $E_F$ must 
be removed and shifted towards the upper and lower Hubbard bands. This is 
precisely the effect induced via the large damping associated with the 
low-frequency resonance in $-$Im\,$\Sigma_X(\omega)$ and the positive 
slope of Re\,$\Sigma_X(\omega)$.

According to this picture, the creation of holes in an electron-doped
Mott insulator also moves the system closer to the insulating phase. 
Thus, spectral weight from states just below $E_F$ must be shifted to 
the Hubbard bands. As discussed in the next subsection, the ED/CDMFT 
results confirm this prediction. The $X$ component of the self-energy 
along the Matsubara axis again exhibits non-Fermi-liquid behavior at 
sufficiently low electron doping. The extrapolation to real $\omega$, 
however, now reveals a resonance slightly below $E_F$, rather than 
above $E_F$ as for hole doping.

\subsection{Electron Doping}

\begin{figure}[t]  
\begin{center}
\includegraphics[width=4.5cm,height=6.5cm,angle=-90]{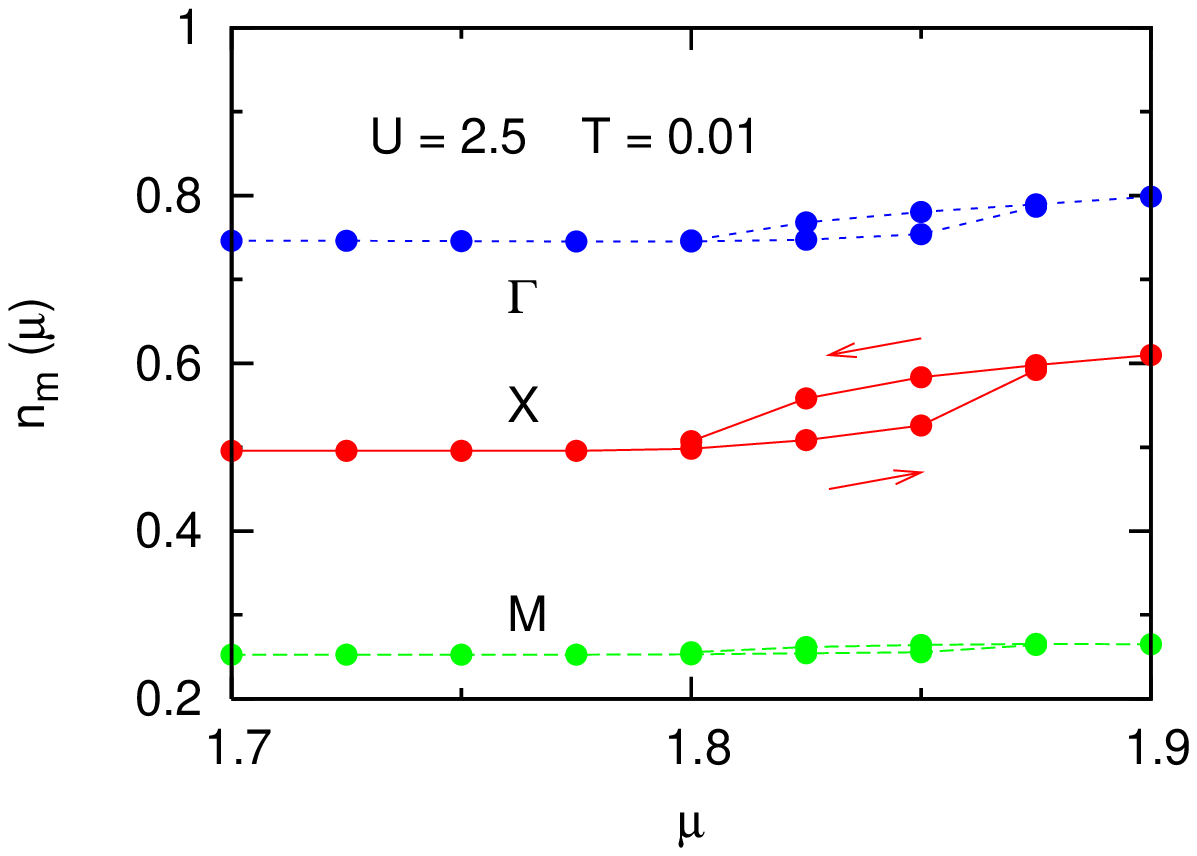}
\includegraphics[width=4.5cm,height=6.5cm,angle=-90]{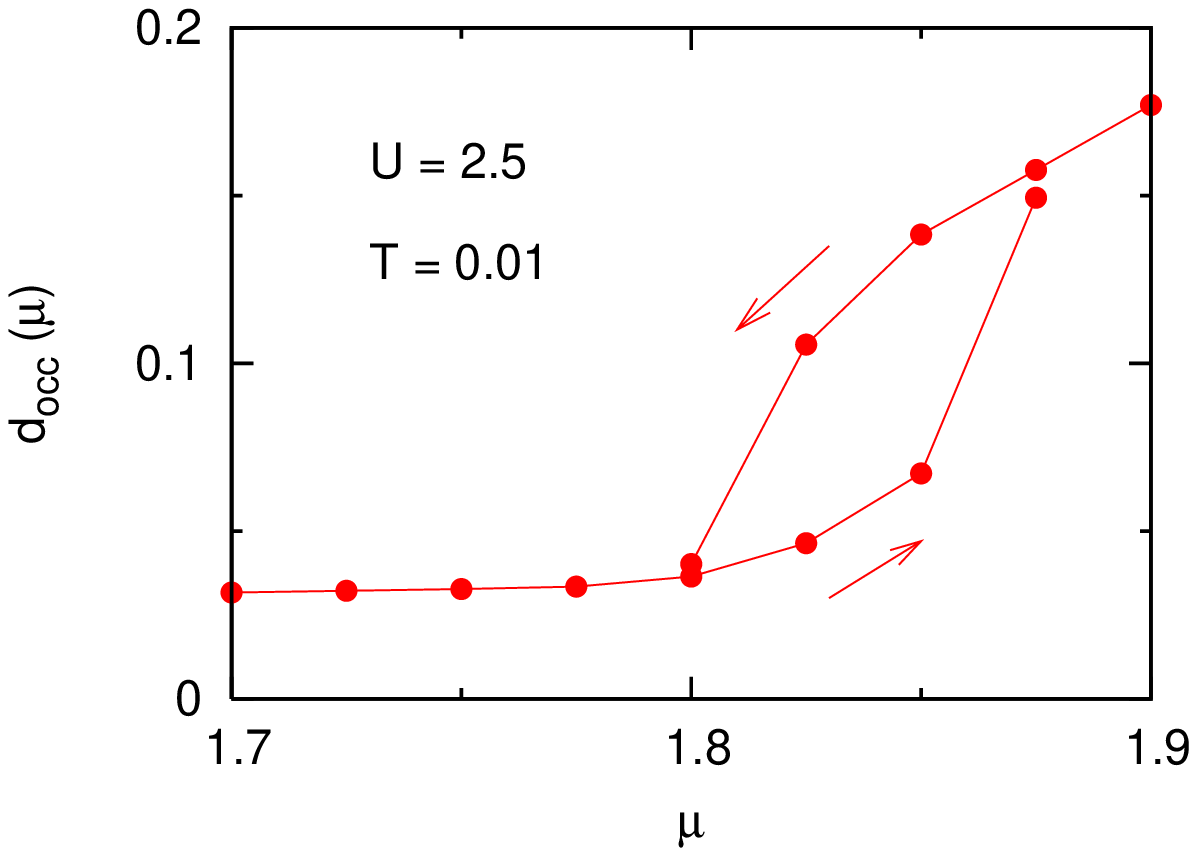}
\end{center}
\caption{(Color online) Upper panel:
Occupancies of cluster molecular orbitals (per spin) as functions of 
chemical potential $\mu$, for $U=2.5$, $T=0.01$. For electron doping,
the Mott transition occurs at about $\mu\approx 1.8\ldots1.875$. In the 
insulating phase $0.7<\mu<1.8$, $n_X=0.5$, $n_\Gamma=0.75$, and $n_M=0.25$
(see also Fig.~\ref{nvsmu}). The arrows denote the hysteresis behavior
for increasing vs.~decreasing $\mu$.
Lower panel: Average double occupancy per site as a function of $\mu$.  
}\label{nvsmu.el}\end{figure}

For completeness we discuss in this subsection the case of electron doping 
which differs from hole doping because of the second-neighbor hopping term 
$t'$. As a result of this interaction, the density of states shown in Fig.~1 
is asymmetric, so that electron doping shifts the van Hove singularity away 
from $E_F$ rather than towards it. Thus, the density of states is reduced 
and less steep. Fig.~\ref{nvsmu.el} shows the occupancies of the cluster 
molecular orbitals in the vicinity of the Mott transition induced via 
electron doping. Both these occupancies as well as the double occupancy
shown in the lower panel exhibit hysteresis behavior for increasing 
vs.~decreasing chemical potential, indicating that this transition is first
order. Thus, this transition is similar to the doping-induced metal insulator
transitions found within local DMFT for single-band and multi-band stystems.
\cite{kotliar:doping,garcia,al:doping}

\begin{figure}[t]  
\begin{center}
\includegraphics[width=4.5cm,height=6.5cm,angle=-90]{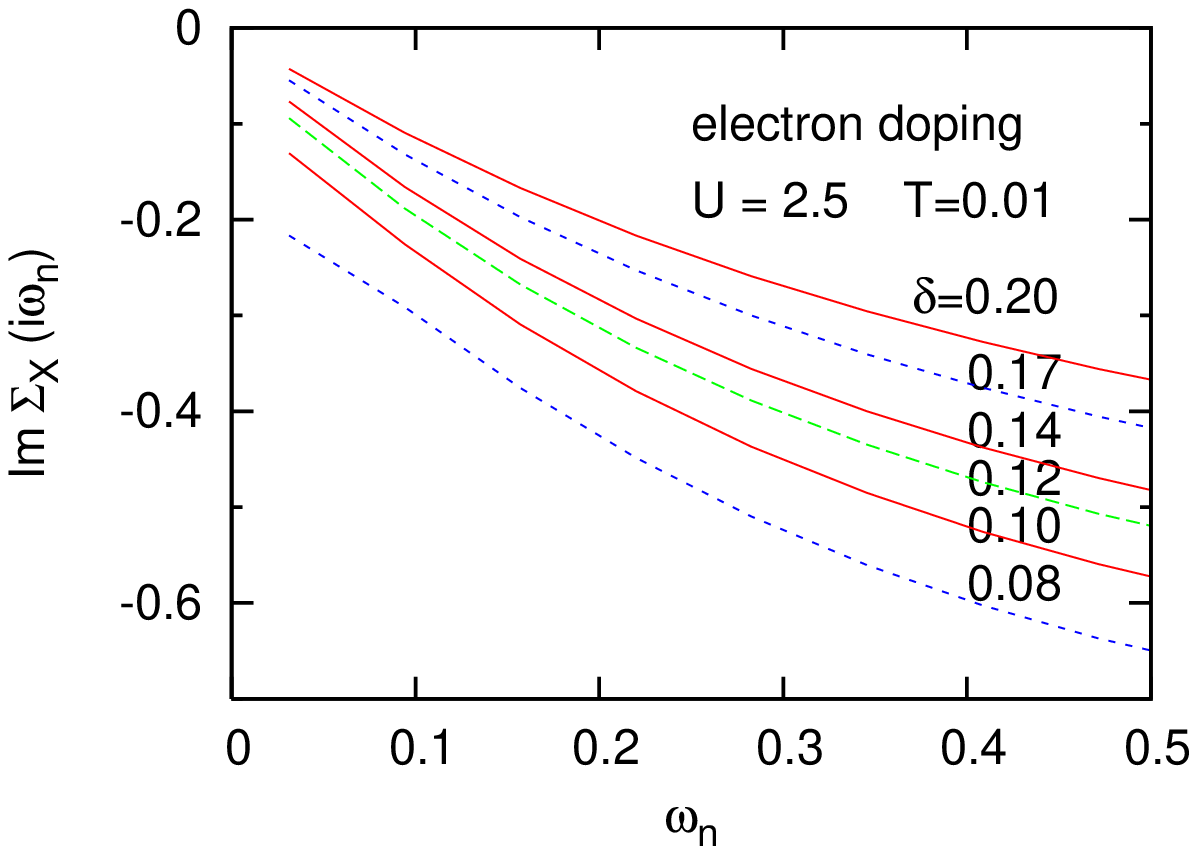}
\includegraphics[width=4.5cm,height=6.5cm,angle=-90]{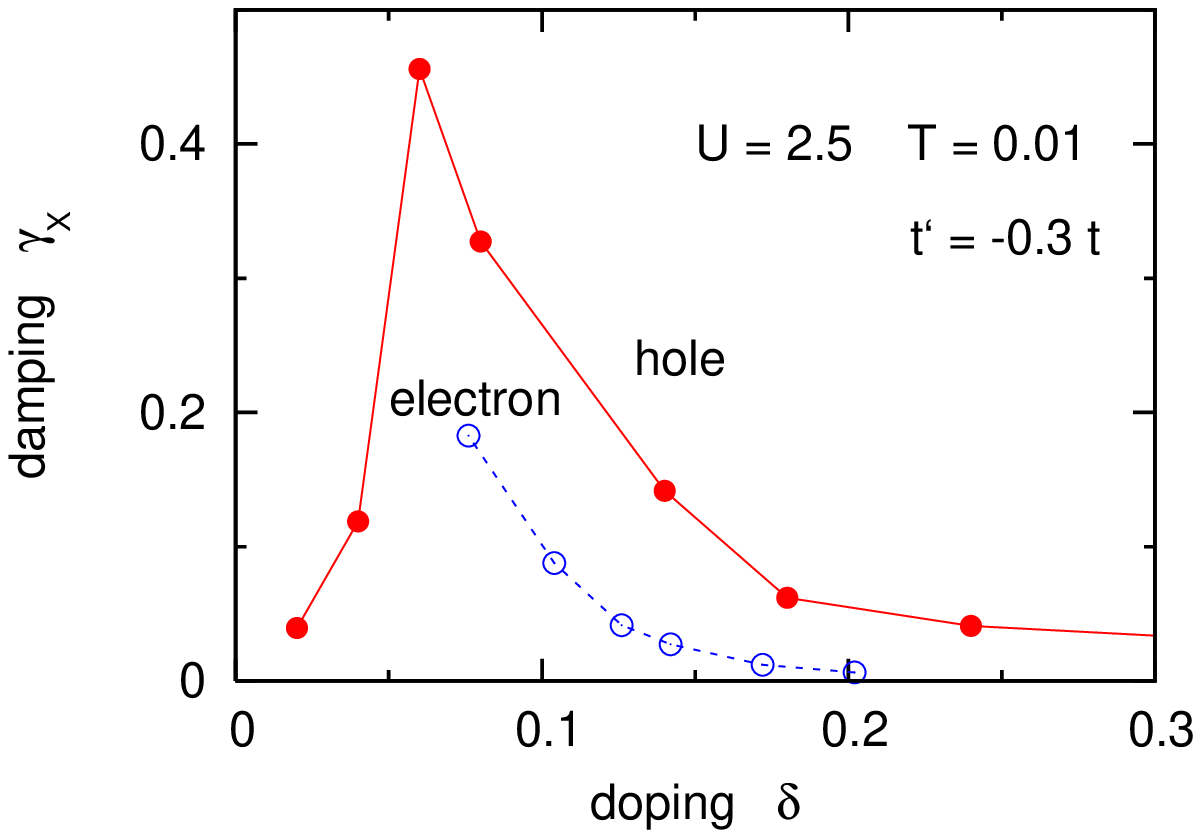}
\end{center}
\caption{(Color online) Upper panel: Imaginary part of $\Sigma_X (i\omega_n)$ 
as a function Matsubara frequency for electron doping; $U=2.5$, $t'=-0.3t$, 
$T=0.01$. The long-dashed green curve for $\delta =0.12$ approximately marks 
the transition from Fermi-liquid to non-Fermi-liquid behavior. The corresponding 
self-energy for hole doping is shown in Fig.~\ref{sigmaT}. 
Lower panel: Comparison of damping 
$\gamma_X= -{\rm Im}\,\Sigma_X (i\omega_n\rightarrow0)$ for hole doping
(solid red circles) and electron doping (empty blue circles).
}\label{eldope}
\end{figure}

Because of the lower and less steep density of states for electron doping,  
the low-frequency variation of the self-energy differs greatly from the hole 
doping case, as illustrated in Fig.~\ref{eldope}. Although there is again a 
clear distinction between Fermi-liquid and non-Fermi-liquid behavior, the 
transition now occurs at considerably smaller $\delta$. While for hole doping 
$\delta_c\approx 0.18\ldots0.20$, for electron doping we find 
$\delta_c\approx 0.12$. Thus, the Fermi-liquid phase is stabilized. 

\begin{figure}[t]  
\begin{center}
\includegraphics[width=4.5cm,height=6.5cm,angle=-90]{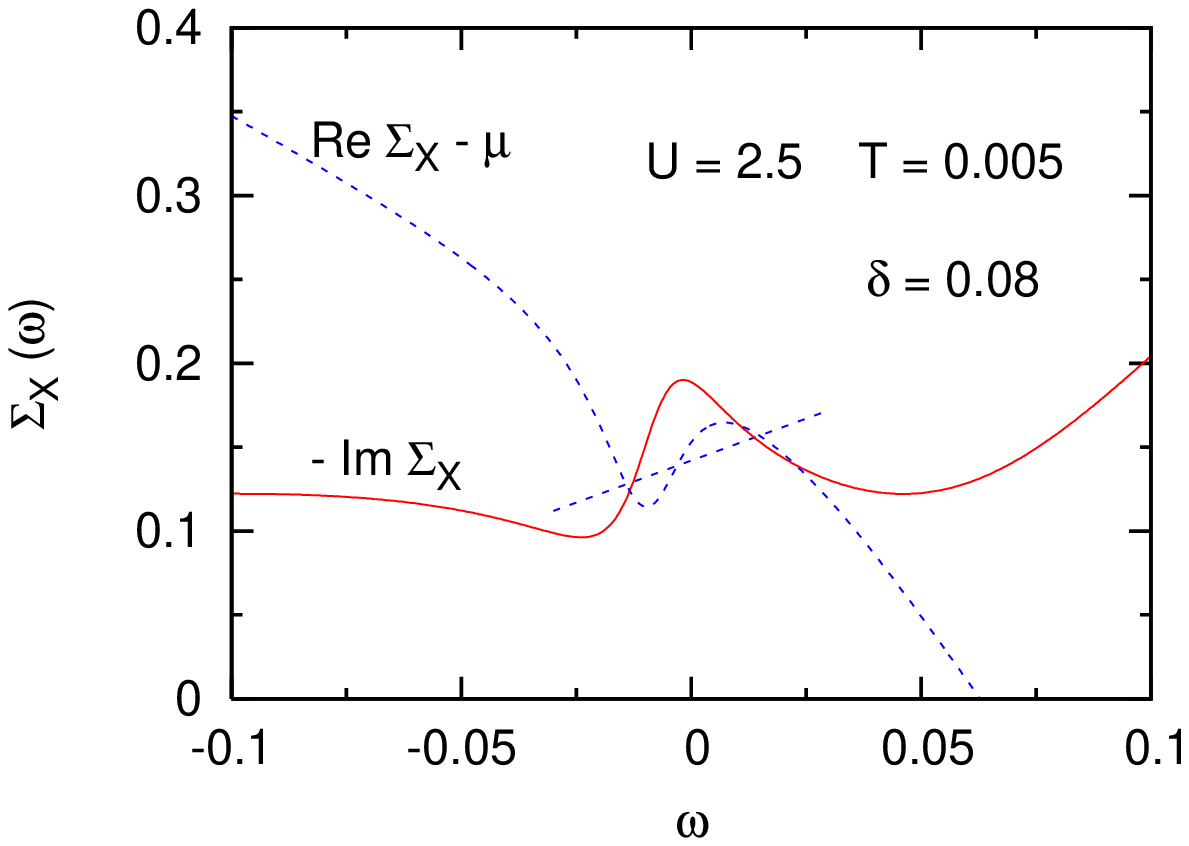}
\includegraphics[width=4.5cm,height=6.5cm,angle=-90]{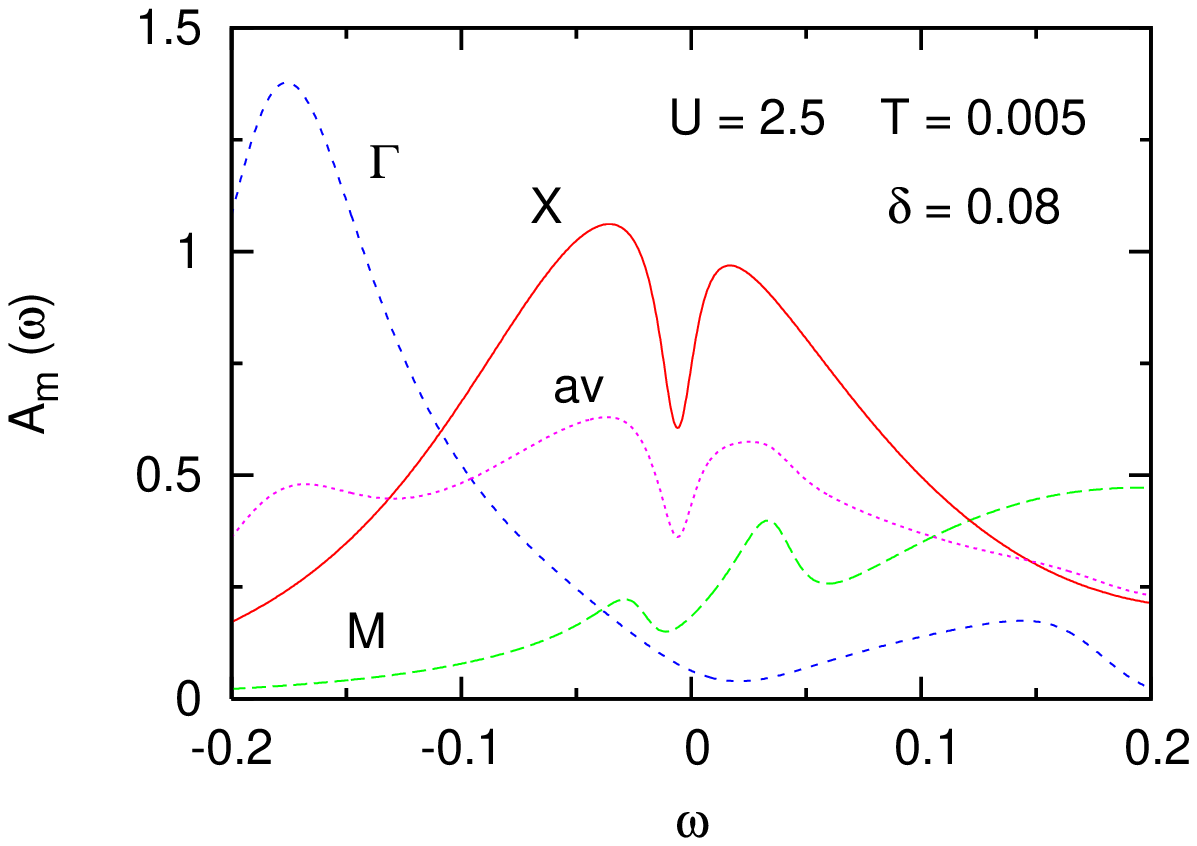}
\end{center}
\caption{(Color online) 
Upper panel: real and imaginary parts of $\Sigma_X(\omega)$ for electron 
doping obtained via extrapolation to real $\omega$. The intersections of 
Re\,$\Sigma_X(\omega)-\mu$ with $\omega-\epsilon_k$ (straight line) 
provide the pseudogap. 
Lower panel: spectral distributions derived via extrapolation of 
lattice Green's function components $G_m(i\omega_n)$ to real $\omega$.
The average density corresponds to 
$A(\omega)=-\frac{1}{4\pi}{\rm Im}\,[G_\Gamma(\omega)+G_M(\omega)+2G_X(\omega)]$. 
}\label{eldopeSigma}
\end{figure}

To identify the pseudogap for electron doping, we evaluate the cluster 
self-energy components via extrapolation to real frequencies. The upper panel 
of Fig.~\ref{eldopeSigma} shows $\Sigma_X(\omega)$ at small $\omega$. 
In this case the non-Fermi-liquid properties give rise to a resonance in 
$-$Im\,$\Sigma_X(\omega)$ centered slightly below the Fermi level, indicating 
that the creation of hole states in an electron-doped Mott insulator implies 
a transfer of spectral weight from states near $E_F$ to the Hubbard bands.
Thereby the system is brought closer to the insulating phase. Accordingly, 
the real part of $\Sigma_X(\omega)$ exhibits a positive slope close to $E_F$. 
Its intersections with $\omega+\mu -\epsilon_k$ can be used to define the 
pseudogap. For $\delta=0.08$ the gap is found to be $\Delta\approx 0.03$, 
i.e., only about half as large as for the hole doping case shown in Fig.~\ref{sr}.   
The lower panel of Fig.~\ref{eldopeSigma} shows the quasiparticle distributions
obtained via extrapolation of the lattice Green's function components, 
Eq.~(\ref{Gm}), to real $\omega$. The dominant feature at small $\omega$ 
is the pseudogap in the $X$ component, which is consistent with the behavior
of $\Sigma(\omega)$ displayed in the upper panel.

The main difference with respect to hole doping, apart from the smaller size
of the pseudogap, is the fact that this gap can be identified only in a very
narrow doping range. At electron doping larger than $0.08$, the non-Fermi-liquid 
behavior is quickly replaced by ordinary Fermi-liquid properties. At smaller
doping, spectral weight just below $E_F$ is rapidly transferred to the 
Hubbard bands, so that the pseudogap is superceded by the opening on the 
Mott gap.

\subsection{Phase Diagram}

\begin{figure}[t]  
\begin{center}
 \includegraphics[width=4.5cm,height=6.5cm,angle=-90]{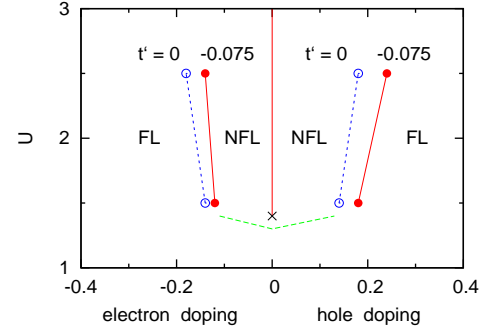}
\end{center}
\caption{(Color online) Phase diagram of two-dimensional Hubbard model
calculated within ED/CDMFT for $2\times2$ clusters. The Fermi-liquid phase
at large hole or electron doping is turned into a non-Fermi-liquid phase at 
small doping. The symbols for $U=1.5$ and $U=2.5$ indicate the approximate 
critical doping $\delta_c$ for $t'=-0.075$ (solid red circles) and for $t'=0$ 
(empty blue circles). The vertical line at $\delta=0$ marks the Mott phase 
at half-filling. The critical $U$ indicated by $\times$ is about $1.4$  
for $t'=0$ and $t'=-0.3t$. The long-dashed green line denotes the 
approximate lower bound of the non-Fermi-liquid domain.
}\label{ph}\end{figure}

In Figure~\ref{Gnud} we have shown that the onset of non-Fermi-liquid
behavior is shifted to smaller hole doping when $U=2.5$ is reduced to 
$U=1.5$ and when $t'=-0.3t$ is replaced by $t'=0$. 
Fig.~\ref{eldope} illustrates the reduction of $\delta_c$ for $U=2.5$ 
when hole doping is replaced by electron doping. A similar reduction
is found for $U=1.5$ (not shown).   
In Fig.~\ref{ph} we collect these data and display the phase diagram 
of the present Hubbard model for electron and hole doping. 
At finite temperature the values of $\delta_c$ can only be determined
within an accuracy of about $\pm0.02$. For clarity, these margins 
are not plotted in Fig.~\ref{ph}. Despite this uncertainty, the 
results demonstrate several trends: for hole doping $\delta_c$ 
diminishes with decreasing $U$ and when $t'=-0.3t$ is replaced by 
$t'=0$. Moreover, for $t'=-0.3t$ the critical doping decreases when 
hole doping is replaced by electron doping. As pointed out above,        
the variation of $\delta_c$ is surprisingly small, despite the 
rather large changes in $U$ and $t'$.

\subsection{Momentum  Variation}

According to the results shown in Fig.~\ref{sigma} the non-Fermi-liquid
properties of the two-dimensional Hubbard model at low hole doping are
mainly associated with the $X$ component of the self-energy. Only very
close to the Mott transition the $M$ component begins to dominate since
its imaginary part changes from $\sim \omega_n$ to $\sim 1/\omega_n$.
The cluster components of the self-energy may be used to construct an
approximate momentum dependent lattice self-energy by using the same
periodization as in Eq.~(\ref{Glat}) for the Green's function. Thus, 
\cite{civelli}
\begin{equation}
\Sigma({\bf k},\omega) =  \alpha_\Gamma({\bf k})\Sigma_\Gamma(\omega)
                         +\alpha_M     ({\bf k})\Sigma_M     (\omega)
                         +\alpha_X     ({\bf k})\Sigma_X     (\omega) 
\label{Slat}
\end{equation}
where
\begin{eqnarray}
 \alpha_\Gamma({\bf k}) &=& [1+{\rm cos} k_x ][1+{\rm cos} k_y ]/4 \nonumber\\
 \alpha_M     ({\bf k}) &=& [1-{\rm cos} k_x ][1-{\rm cos} k_y ]/4          \\
 \alpha_X     ({\bf k}) &=& [1-{\rm cos} k_x     {\rm cos} k_y ]/2 .\nonumber         
\end{eqnarray}
The ${\bf k}$-resolved spectral distributions are then given by 
\begin{equation}
    A({\bf k},\omega)= -\frac{1}{\pi}{\rm Im}\,(\omega+\mu-\epsilon({\bf k})
            -\Sigma({\bf k},\omega))^{-1} .
\label{Ako}
\end{equation}
 
An alternative is to periodize instead the cumulant matrix\cite{stanescu2} 
$M(\omega) = 1/[\omega+\mu-\Sigma(\omega)]$
which can be diagonalized in the same manner as the self-energy. 
Thus, the molecular orbital components of $M(\omega)$ are given by       
$M_m(\omega) = 1/[\omega+\mu-\Sigma_m(\omega)]$ and the momentum 
dependent lattice cumulant $M({\bf k},\omega)$ can be derived 
from an expression analogous to Eq.~(\ref{Slat})     
\begin{equation}
M({\bf k},\omega) =  \alpha_\Gamma({\bf k})M_\Gamma(\omega)
                         +\alpha_M({\bf k})M_M(\omega)
                         +\alpha_X({\bf k})M_X(\omega) .
\label{Mlat}
\end{equation}
The lattice self-energy in this approximation takes the form 
\begin{equation}
 \Sigma({\bf k},\omega)= \omega+\mu- 1/M({\bf k},\omega) .
\label{SlatM}
\end{equation}

\begin{figure}[t]  
\begin{center}
 \includegraphics[width=4.5cm,height=6.5cm,angle=-90]{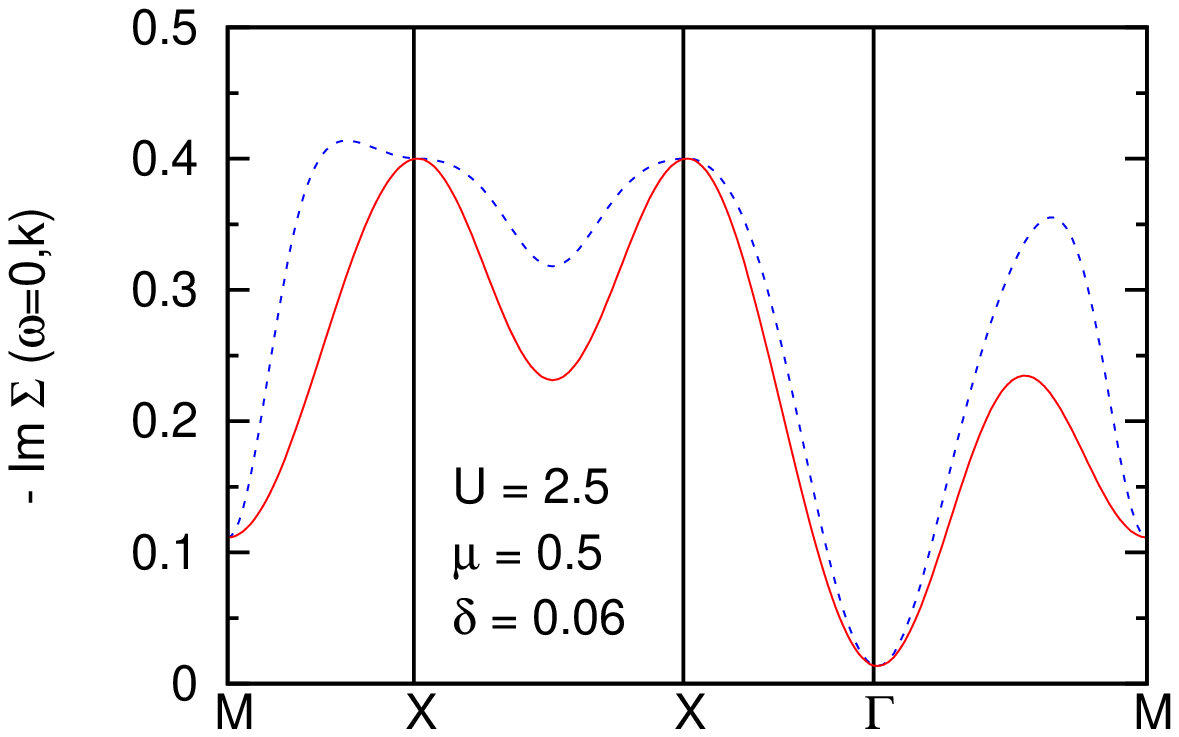}
 \includegraphics[width=4.5cm,height=6.5cm,angle=-90]{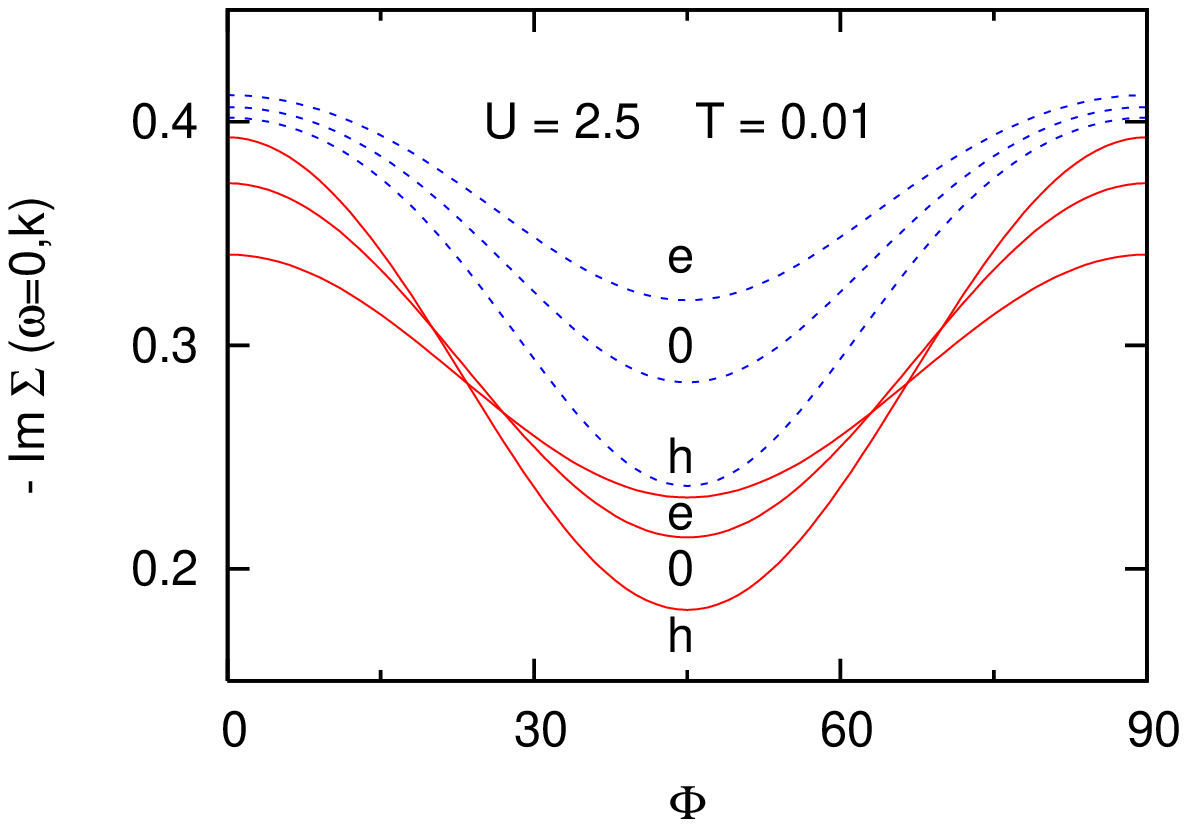}
\end{center}
\caption{(Color online) 
Upper panel: momentum variation of imaginary part of $- \Sigma_X(\omega,{\bf k})$ 
at Fermi level. 
Solid red curve: expression Eq.~(\ref{Slat}), corresponding to 
periodization of cluster self-energy; dashed blue curve: expression
Eq.~(\ref{SlatM}), corresponding to periodization of cumulant;
$U=2.5$, $\delta=0.06$, $T=0.01$.
Lower panel: azimuthal variation of $-$Im\,$\Sigma_X(\omega=0,{\bf k})$  
for ${\bf k}=\pi(1-r{\rm sin}\Phi,1-r{\rm cos}\Phi)$ with $r=0.7,\ 0.8,\ 0.9$,
approximately representing electron doping (e), half-filling (0), and 
hole doping (h), respectively. $\Phi=0, 90$ corresponds to ${\bf k}$ along 
$XM$ and  $\Phi=45$ to the nodal direction $\Gamma M$. 
Solid red curves: periodization via Eq.~(\ref{Slat}); 
dashed blue curves: periodization via Eq.~(\ref{SlatM}).   
}\label{disp}\end{figure}

In the upper panel of Fig.~\ref{disp} we compare these two versions of 
$\Sigma({\bf k},\omega)$
at $\omega=0$ for $\delta=0.06$ hole doping. The real-$\omega$ components
$\Sigma_m(\omega=0)$ are obtained via extrapolation from the first few
Matsubara frequencies. At high-symmetry points both versions of 
Im\,$\Sigma({\bf k},\omega=0)$ coincide. At general ${\bf k}$-points, 
however, the cumulant expression yields enhanced damping, in particular, 
between $M$ and $X$, and along $\Gamma M$. The enhancement near $X$ leads 
to an effective flattening of Im\,$\Sigma({\bf k},\omega)$, which is also 
seen in the dual Fermion approach.\cite{rubtsov} On the other hand, it is 
not clear whether this enhancement is partly an artifact of the cumulant 
approximation since the damping at some points between $X$ and $M$ is 
even larger than at $X$. 
Also, damping near ${\bf k}\approx 2/3 (\pi,\pi)$ in the cumulant version 
is almost as large as at $X$. At the present doping ($\delta=0.06$), the 
periodization of the self-energy according to Eq.~(\ref{Slat}) is in better 
agreement with the dual Fermion approach (see Fig.~15 of Ref.~\cite{rubtsov}).     

The lower panel of Fig.~\ref{disp} shows the variation of 
$-$Im\,$\Sigma(\omega=0,{\bf k})$ along  
${\bf k}=\pi(1-r{\rm sin}\Phi,1-r{\rm cos}\Phi)$ where $r=0.7,\ 0.8,\ 0.9$
is chosen to approximately represent the Fermi arcs for electron doping, 
half-filling, and hole doping, respectively. Both periodization versions
yield consistently larger damping along $XM$ than along the nodal direction 
$\Gamma M$. The cumulant version implies overall larger damping, and, more 
importantly, less pronounced difference between $\Gamma M$ and $XM$. 
Because of the substantial imaginary part of the self-energy at low
frequencies, the Fermi surface in the present $2\times2$ cluster approach
exhibits arcs rather than hole pockets.\cite{civelli} We note, however, 
that greater momentum differentiation obtained for larger clusters might 
lead to more pronounced anisotropy between the nodal and anti-nodal 
directions. In particular, this could yield smaller values of 
$-$Im\,$\Sigma(\omega=0,{\bf k})$ along $\Gamma M$ ($\Phi=45$) than 
indicated in Fig.~\ref{disp}.

\begin{figure}[t]  
\begin{center}
 \includegraphics[width=4.5cm,height=6.5cm,angle=-90]{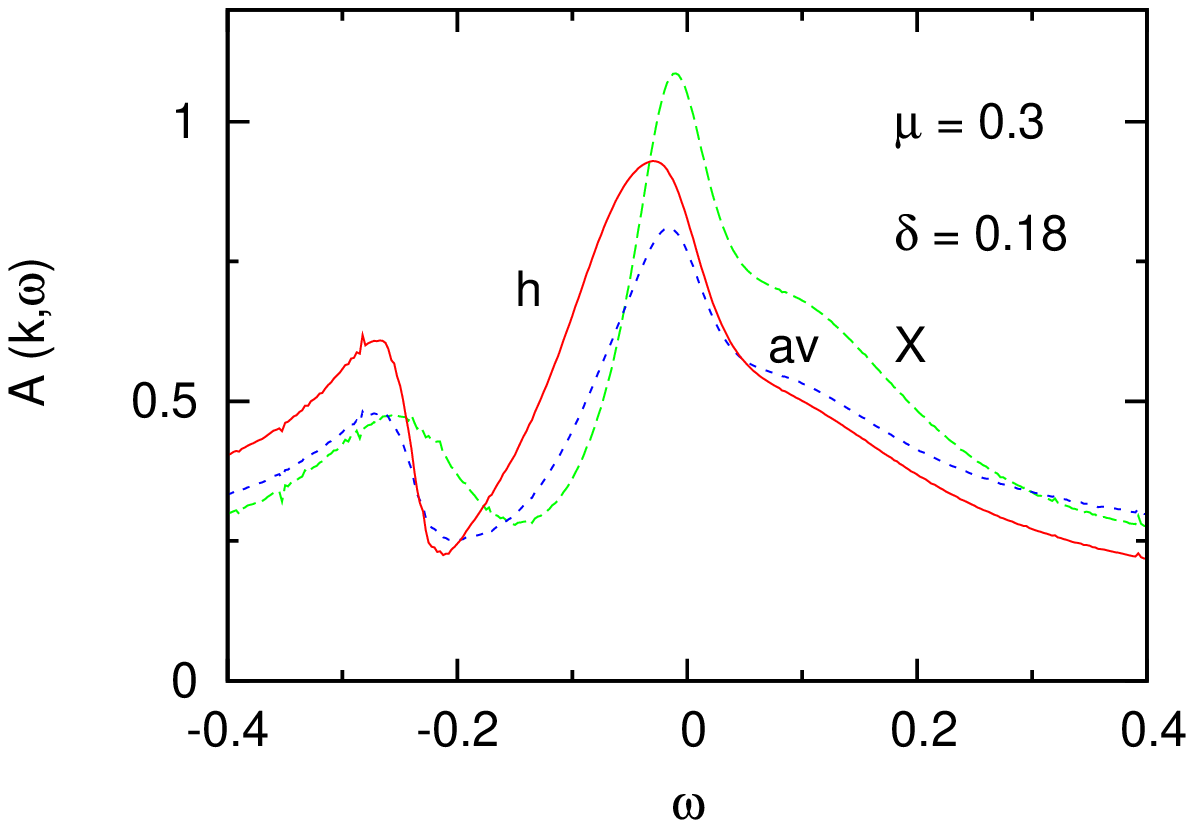}
 \includegraphics[width=4.5cm,height=6.5cm,angle=-90]{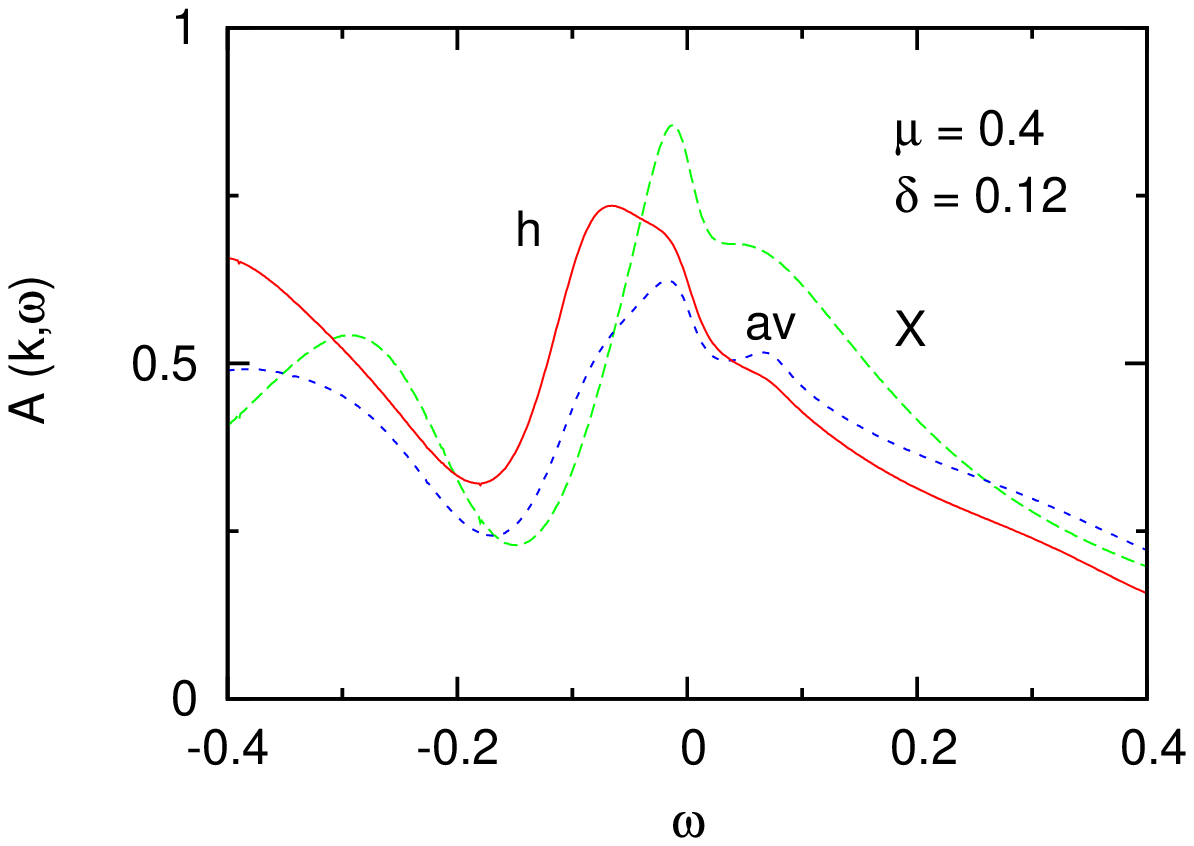}
 \includegraphics[width=4.5cm,height=6.5cm,angle=-90]{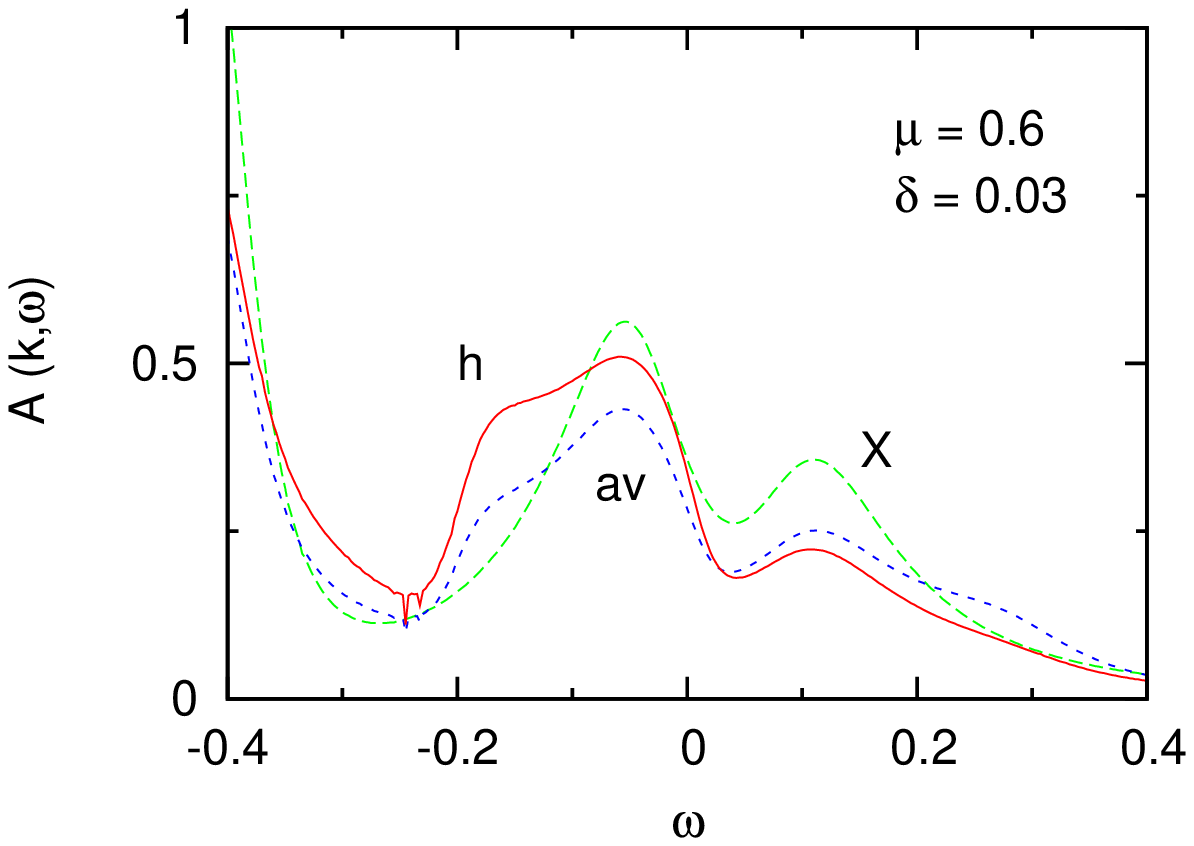}
\end{center}
\caption{(Color online) 
Lattice spectral distributions $A({\bf k},\omega)$ for three chemical
potentials close to critical doping ($\mu=0.3$, upper panel), incipient
pseudogap formation  ($\mu=0.4$, middle panel), and complete pseudogap
phase  ($\mu=0.6$, lower panel).    
Solid red curves ($h$):  $k_x=k_y=0.36\pi$ near Fermi surface for hole 
doping; dashed blue curves ($av$):  $k_x=k_y=0.5\pi$ at center of Brillouin 
zone, corresponding to average density; long-dashed green curves ($X$):  
${\bf k}=(\pi,0)=X$; $U=2.5$, $T=0.01$.    
}\label{A.mu}\end{figure}

\begin{figure}[t]  
\begin{center}
\includegraphics[width=6.5cm,height=8.5cm,angle=-90]{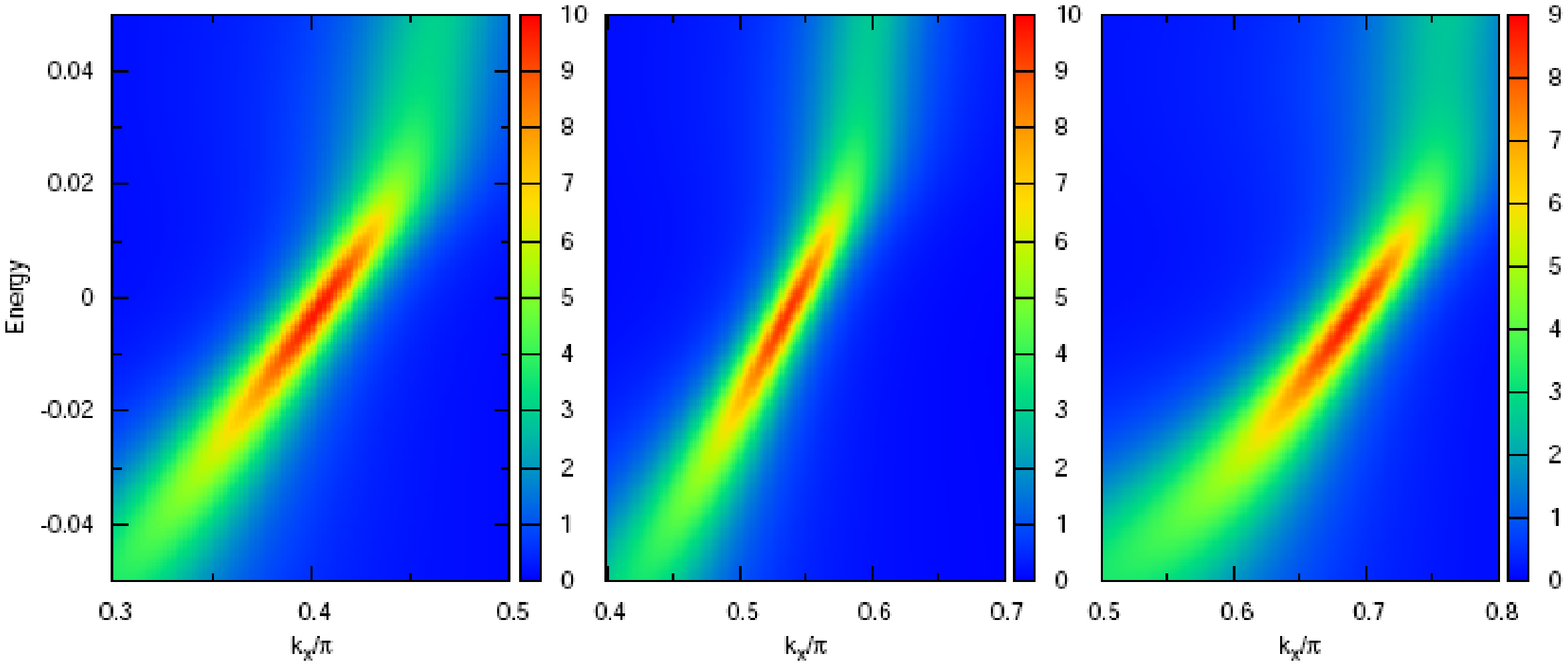}
\includegraphics[width=6.5cm,height=8.5cm,angle=-90]{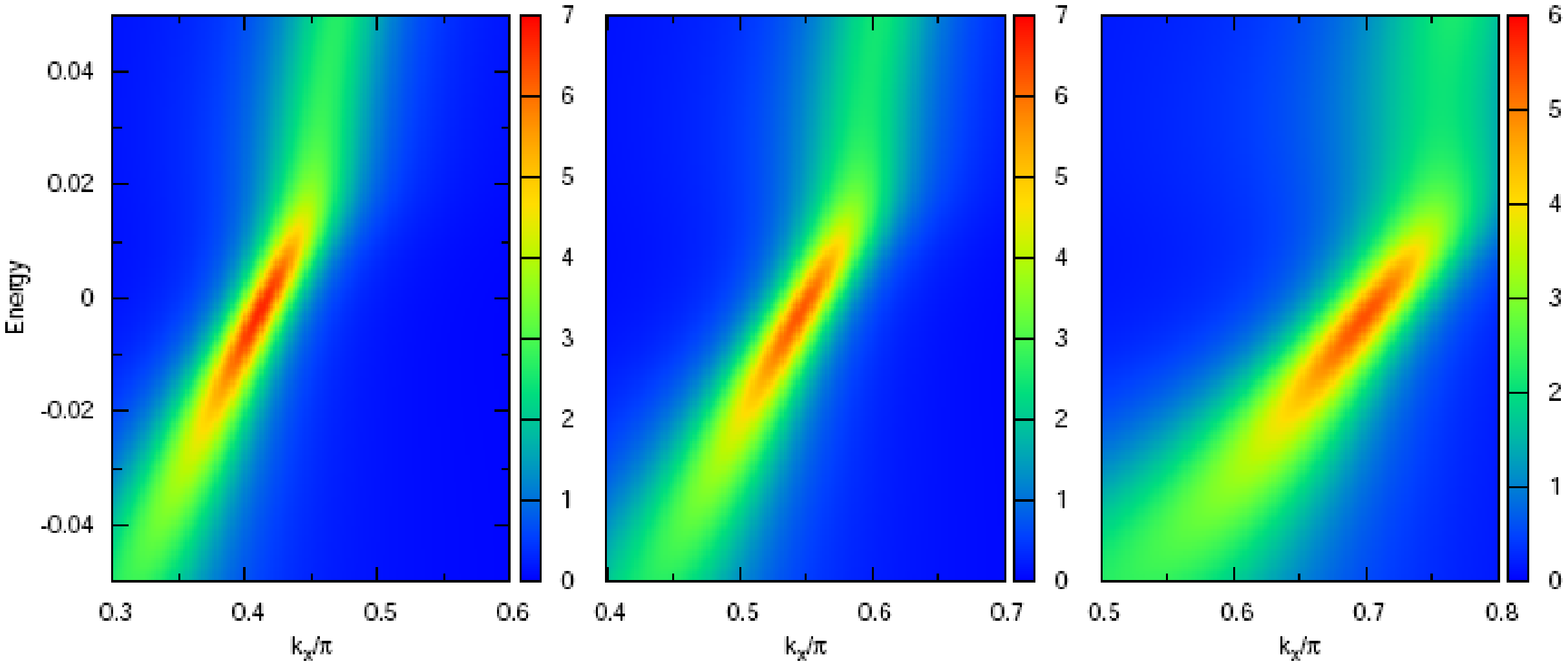}
\includegraphics[width=3.0cm,height=3.0cm,angle=-90]{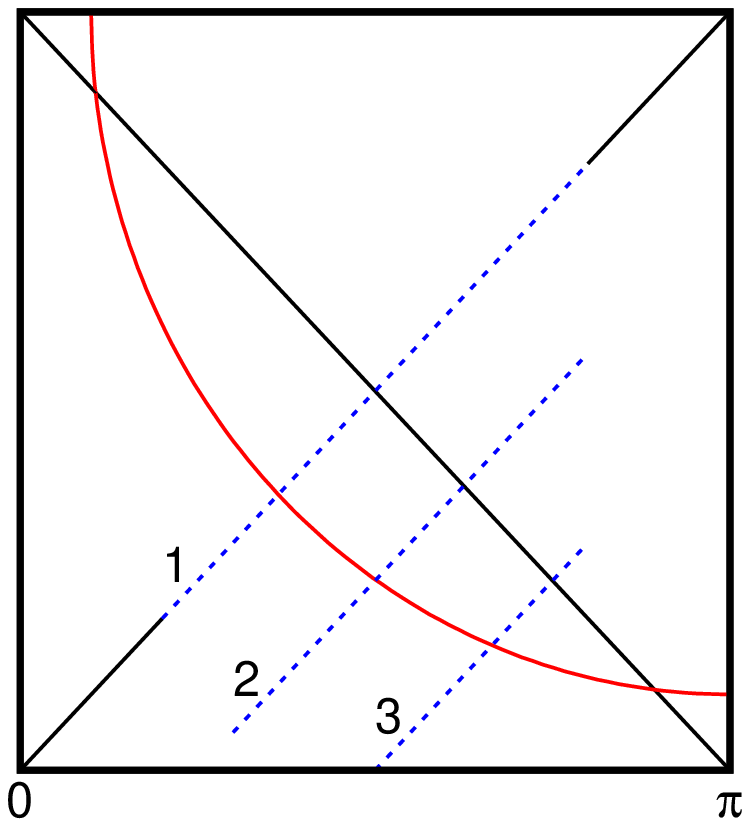}
\end{center}
\caption{(Color online) 
Upper three panels:
Spectral function $A({\bf k},\omega)$ along cuts 1, 2, 3 (from left to right);
top panel: doping $\delta = 0.17$; middle panel:  $\delta = 0.14$; 
$U=2.5$, $T=0.005$.
Lower panel: 
Cuts through Brillouin zone corresponding to ARPES data in Ref.\cite{yang3}.
Solid red curve: approximate non-interacting Fermi surface for hole doping. 
}\label{a.mu1} \end{figure}

\begin{figure}[t]  
\begin{center}
\includegraphics[width=6.5cm,height=8.5cm,angle=-90]{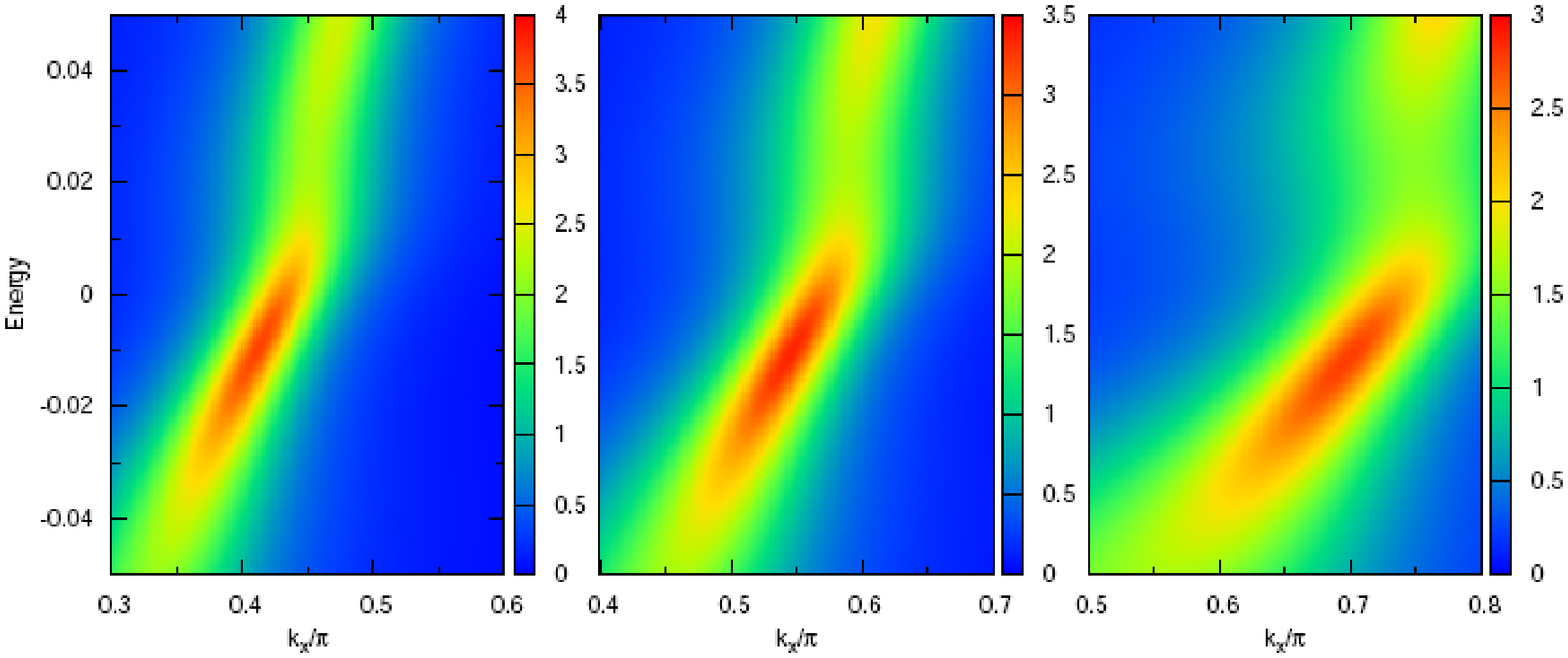}
\includegraphics[width=6.5cm,height=8.5cm,angle=-90]{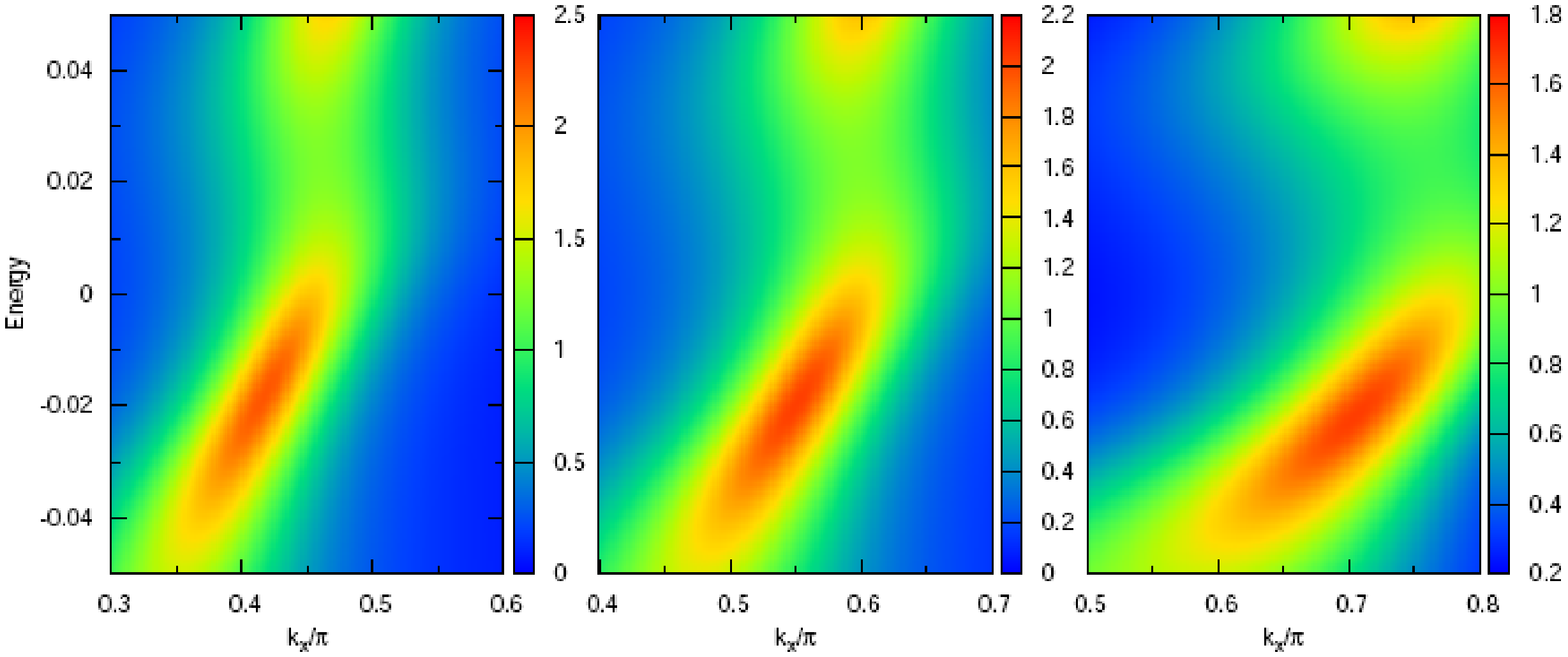}
\end{center}
\caption{(Color online) 
Same as Fig.~\ref{a.mu1} except  $\delta = 0.11$ (upper panel) 
and $\delta = 0.08$ (lower panel). 
}\label{a.mu2}\end{figure}

According to the collective mode in Im\,$\Sigma_X(\omega=0)$ (see Fig.~\ref{sr}),
the anisotropy between the $\Gamma X$ and $\Gamma M$ directions is even 
larger above $E_F$ than for the $\omega\rightarrow0$ limit shown
in Fig.~\ref{disp}. Thus, the collective mode gives rise to a momentum 
and doping dependent particle-hole asymmetry. To illustrate this point,
we show in Fig.~\ref{A.mu} the low-frequency part of the spectral
distribution $A({\bf k},\omega)$, derived via extrapolation of the 
lattice Green's function, Eq.~(\ref{Glat}), at three representative 
points in the Brillouin zone. Three doping regions
can be distinguished: At $\mu=0.3$ close to optimal doping ($\delta=0.18$, 
upper panel), there is weak anisotropy since the system 
is a Fermi liquid throughout ${\bf k}$ space. Below critical doping
($\mu=0.4$, $\delta=0.12$, middle panel), the spectrum in the anti-nodal
direction at $X$ shows clear signs of pseudogap behavior, while the one at 
${\bf k}=0.36(\pi,\pi)$, i.e., near the nodal point of the Fermi surface 
for hole hoping, is still dominated by Fermi-liquid properties. At this 
${\bf k}$ point, the coefficients in the momentum expansion, Eq.~(\ref{Slat}),
are $(\alpha_\Gamma,\alpha_M,\alpha_X)=(0.50, 0.09,0.41)$, indicating the 
rather large Fermi-liquid-like $\Gamma$ component. At the zone center 
these coefficients are $(1/4,1/4,1/2)$. Finally, at even lower doping
($\mu=0.6$, $\delta=0.03$, bottom panel), close to the Mott transition,
the non-Fermi-liquid properties have spread across the entire Fermi 
surface, so that the pseudogap is observable along the nodal as well 
as antinodal directions. These results demonstrate the non-uniform,
momentum dependent opening of the pseudogap as a function of doping.
(Note that this behavior differs from the opening of the Mott gap shown 
in Fig.~\ref{Am}, which in the present $2\times2$ cluster DMFT takes 
place simultaneously in all cluster components.)



To analyze the particle-hole asymmetry observed in the recent ARPES data 
on Bi$_2$Sr$_2$CaCu$_2$O$_{8+\delta}$ by Yang {\it et al.}\cite{yang3} 
we have calculated the spectral distributions $A({\bf k},\omega)$ defined
in Eq.~(\ref{Ako}), where the self-energy is obtained from Eq.~(\ref{Slat}). 
The frequency variation of the $X$ component is shown in Fig.~\ref{sr}.
For direct comparison with the data we plot $A({\bf k},\omega)$ along 
three cuts, as indicated in the lower panel of Fig.~\ref{a.mu1}. 
Cut 1 corresponds to the nodal direction and has the lowest relative 
weight from $\Sigma_X(\omega)$, while in cut 3 the $X$ component dominates.  

For large doping (Fig.~\ref{a.mu1}, top panel), the system is a Fermi liquid. 
Thus, the spectral weight at all three cuts is largest at $E_F$ and decays
symmetrically for increasing and decreasing $\omega$. 
Below critical doping (middle panel), this particle-hole symmetry 
initially persists along the nodal direction, but gets weaker along cut 3.
At $\delta=0.08$  (Fig.~\ref{a.mu2}, upper panel), this asymmetry begins to 
extend to the nodal direction, until at $\delta=0.08$  (lower panel) the 
particle-hole asymmetry is complete throughout the Brillouin zone. 
These spectral distributions reveal that the particle-hole asymmetry 
is a direct consequence of the pseudogap which gradually develops with
doping at about $0.02\ldots0.05$ above $E_F$, and which is driven by 
the $(\pi,0)$ component of the self-energy.      
  
The momentum dependent opening of the pseudogap above $E_F$, and the 
particle-hole asymmetry caused by the collective mode seen in 
Im\,$\Sigma_X(\omega)$ (see Fig.~\ref{sr}), are in excellent agreement 
with the ARPES data.\cite{yang3} Although DMFT calculations 
for even larger clusters provide even better momentum differentiation,
the present results for $2\times2$ clusters reveal that spatial
degrees of freedom give rise to dramatic new phenomena absent in a local
description, in particular, the resonance in the $(\pi,0)$ component 
of the self-energy at small positive frequencies. It would be very 
interesting to check whether the dispersion of the position of this 
resonance with doping can be verified experimentally.

\section{Summary}

The effect of short-range correlations in the two-dimensional Hubbard model
is studied within finite-temperature ED combined with DMFT for $2\times2$ 
clusters. A mixed basis consisting of cluster sites and bath molecular 
orbitals is shown to provide an efficient and accurate projection of the 
lattice Green's function onto the cluster. The onset of non-Fermi-liquid 
behavior with decreasing hole doping is evaluated for various Coulomb
energies, temperatures, and next-nearest neighbor hopping interactions.
The self-energy component $\Sigma_{X=(\pi,0)}(\omega)$ is shown to 
exhibit a collective mode above $E_F$ which becomes more intense close
to the Mott transition. This resonance implies the removal of spectral 
weight from electron states above to $E_F$ and the opening of a pseudogap. 
With decreasing doping the pseudogap opens first along the antinodal 
direction and then spreads across the entire Fermi surface.
For electron doping, the resonance of $\Sigma_X(\omega)$ and the 
corresponding pseudogap are located below $E_F$, as expected for the 
removal of hole states close to $E_F$. In the low doping range 
the density of states at the Fermi level becomes very asymmetric. 
Near the onset of non-Fermi-liquid behavior, $E_F$ is at a maximum of the 
density of states. At smaller doping $E_F$ moves into the pseudogap. 
This behavior leads to a pronounced particle-hole asymmetry in 
the spectral distribution at intermediate hole doping, in agreement
with recent ARPES measurements. The phase 
diagram shows that for hole doping $\delta_c\approx 0.15\ldots0.20$ for 
various system parameters, i.e., near the optimal doping observed in 
many high-$T_c$ cuprates. The critical electron doping which marks the 
onset of non-Fermi-liquid behavior is systematically smaller than for 
hole doping. The Mott transition induced via electron doping exhibits
first-order hysteresis characteristics. In contrast, within the present 
cluster ED/DMFT the hole doping transition appears to be continuous or
weakly first-order at very low temperatures. 

\bigskip

{\bf Acknowledgements}\ \ 
N.-H. T. is supported by the Alexander von Humboldt  Foundation. 
The computational work was carried out on the J\"ulich JUMP.

\end{document}